\documentclass[showpacs,amssymb,floatfix]{revtex4}
\usepackage{graphicx}
\usepackage{dcolumn}
\usepackage{graphics}
\usepackage{epstopdf}
\usepackage{pdfpages}
\usepackage{bm}
\begin{document}
\title{Light trapping above the light cone in one-dimensional array of dielectric spheres}
\author{Evgeny N. Bulgakov$^{1,2}$ and Almas F. Sadreev$^1$}
\affiliation{$^1$ L.V. Kirensky Institute of Physics, 660036 Krasnoyarsk,
Russia\\
$^2$Siberian State Aerospace University, Krasnoyarsk 660014, Russia}
\date{\today}
\begin{abstract}
We demonstrate bound states in the first TE and TM diffraction continua (BSC) in a linear periodic  array of dielectric spheres
in air above the light cone.
We classify the BSCs according to the symmetry specified by the azimuthal number $m$, the Bloch wave vector $\beta$
directed along the array, and polarization. The most simple symmetry protected TE and TM polarized BSCs
have $m=0$ and  $\beta=0$ and occur in a wide range of the radius of the spheres and dielectric constant. More
complicated BSCs with $m\neq 0$ and $\beta=0$  exist
only for a selected radius of spheres at a fixed dielectric constant. We also find robust Bloch BSCs with
$\beta\neq 0, m=0$.
We present also the BSCs embedded into two and three diffraction continua.
We show that the BSCs can be easily detected by the collapse of Fano resonance for scattering of
electromagnetic plane waves
by the array.
\end{abstract}
 \maketitle
%%%%%%%%%%%%%%%%%%%%%%%%%%%%%%%%%%%%%%%%%%%%%%%%%%%%%%%%%%%%%%%%%%%%%%%%%%%%%%%%%%%%%%%%%%%%%%%%%%%%%%
\section{Introduction}
%%%%%%%%%%%%%%%%%%%%%%%%%%%%%%%%%%%%%%%%%%%%%%%%%%%%%%%%%%%%%%%%%%%%%%%%%%%%%%%%%%%%%%%%%%%%%%%%%%%%%%%%
The scattering of electromagnetic (EM) waves by an ensemble of dielectric spheres
has a long history of research  beginning with Mie who presented a rigorous theory
for scattering by a single dielectric sphere \cite{Stratton}.
The overwhelming majority of papers since the pioneering papers by
Ohtaka and his coauthors \cite{Ohtaka,OhtakaJPC,Inoue,Miyazaki} have
considered periodical two-dimensional (2D) and three-dimensional (3D) arrays \cite{Modinos,Bruning,Abajo,Wang}.
Surprisingly, less interest was payed to scattering by a linear
array of dielectric spheres mostly restricted to aggregates of
a finite number of spheres \cite{FullerI,Hamid,Xu}.
Guiding of electromagnetic waves by a linear array of dielectric spheres below the diffraction
limit attracted more attention.
There were two types of consideration: finite arrays \cite{Mackowski,Quinten,Quirantes,Luan,Zhao,Du1}
and infinite arrays which were studied by means of the coupled-dipole approximation
\cite{Gozman,Blaustein,Draine,Savelev,Krasnok}.  A consummate analysis of
electromagnetic waves  propagating along linear arrays of
dielectric spheres below the light cone
 was provided by Linton {\it et al}
\cite{Linton_Zal}.

It has been widely believed that only
those modes whose eigenfrequencies lie below the light cone, are confined and the
rest of the eigenmodes have finite life times. Recently confined electromagnetic modes were shown to exist
in various periodical arrays of (i) long cylindrical rods \cite{Shipman,Ndangali,PRA,Hu&Lu},
(ii) photonic crystal slabs \cite{Wei,Wei1,Yang,Bo Zhen}, and (iii) 2D arrays of spheres
\cite{Zhang,Song}. Similarly one may expect light trapping in
the one-dimensional (1D) array of spheres with the bound frequencies
{\it above the light cone}. Such localized solutions of the Maxwell
equations are known as the bound states in the continuum (BSC) and
were first reported by von Neumann and Wigner \cite{neumann} for the stationary Schr\"odinger
equation with a specially chosen oscillatory potential. Nowadays,
the BSCs are known to exist in various waveguide structures
ranging from quantum dots \cite{Nockel,Kim, Olendski,
SBR}, to acoustic periodic structures \cite{Porter0,Porter1,Porter,Linton,Colquitt} where they are known
as the embedded trapped modes, to photonic crystals
\cite{Yang,photonic,Shabanov,Longhi1,Longhi2,robust,Rivera}. The BSCs are of
immense interest in optics thanks to the experimental
opportunity to confine light despite that
outgoing waves are allowed in the surrounding medium \cite{Bo Zhen,Plotnik,Lopez,Longhi,Kivshar,Song}.

%%%%%%%%%%%%%%%%%%%%%%%%%%%%%%%%%%%%%%%%%%%%%%%%%%%%%%%%%%%%%%%%%%%%%%%%%%%%%%%%%%%%%%%%%%%%%%%
\section{Basic equations for electromagnetic wave scattering by a linear array of spheres}
\begin{figure}[ht]
\includegraphics[width=7cm,clip=]{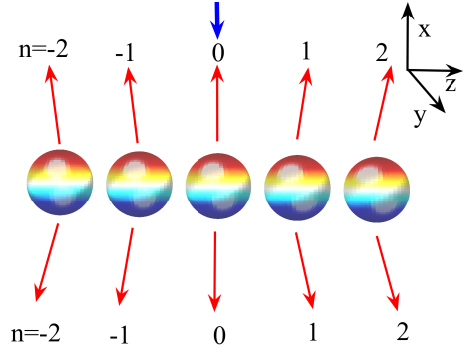}
\caption{A periodic infinite array of dielectric spheres illuminated by a plane
wave (blue short arrow). The wave can be transmitted and reflected to discrete
diffraction continua enumerated by integers $m$ and $n$ in accordance with Eqs. (\ref{psi}) and (\ref{kz})
shown by red long arrows.}
\label{fig1}
\end{figure}

%%%%%%%%%%%%%%%%%%%%%%%%%%%%%%%%%%%%%%%%%%%%%%%%%%%%%%%%%%%%%%%%%%%%%%%%%%%%%%%%%%%%%%%%%%%%%%%%
In the present paper we consider a
free-standing 1D infinite array of dielectric spheres
in air Fig. \ref{fig1}.
In what follows we refer to all length quantities in terms of the period $h$ of the array.
%-------------------------------------------------------------------------------------------------------Fig.1
We formulate the scattering theory by a periodic array of dielectric
spheres in the form of the Lippmann-Schwinger equation similar to the approach developed for
a periodic array of dielectric cylinders \cite{PRA}
%---------------------------------------------------------------------------------------------(1)
\begin{equation}\label{LS}
\widehat{L}\mathbf{a}=\mathbf{\Psi}_{inc}.
\end{equation}
where the matrix $\widehat{L}$ accounts for both
the scattering matrix of the isolated sphere and the mutual scattering events between the
spheres, $\mathbf{\Psi}_{inc}$ is given by the incident wave, and the column $\mathbf{a}$ consists of amplitudes
$a_l^m$ of the multipole expansion of the scattering function.

The exact expression of the matrix $\widehat{L}$ was derived by Linton {\it et al} \cite{Linton_Zal}
for EM guided waves on a periodic array of dielectric spheres.
For the reader's convenience we present the equations and notations from the above reference.
We seek the solutions of the Maxwell equations, which obey the Bloch theorem
$$\mathbf{E}(\mathbf{r}+\mathbf{R}_j)=e^{i\mathbf{\beta}\mathbf{R}_j}\mathbf{E}(\mathbf{r}),
\mathbf{H}(\mathbf{r}+\mathbf{R}_j)=e^{i\mathbf{\beta}\mathbf{R}_j}\mathbf{H}(\mathbf{r})$$
with the Bloch wave vector $\mathbf{\beta}$ directed along the array aligned with the z-axis (see Fig. \ref{fig1}).
Here $\mathbf{R}_j=j\mathbf{e}_z$ is
the position of the center of the j-th sphere and $\mathbf{e}_z$ is the unit vector along the array. Scattered
EM fields are expanded
in a series over vector spherical harmonics $\mathbf{M}_n^m$ and $\mathbf{N}_n^m$ \cite{Stratton,Linton_Zal} defined in Appendix A
%--------------------------------------------------------------------------------------------------(2)
\begin{eqnarray}\label{out}
&\mathbf{E}(\mathbf{r})=\sum_je^{i\mathbf{\beta}\mathbf{R}_j}\sum_{lm}[a_l^m
\mathbf{M}_l^m(\mathbf{r}-\mathbf{R}_j)+
b_l^m\mathbf{N}_l^m(\mathbf{r}-\mathbf{R}_j)],&\nonumber\\
&\mathbf{H}(\mathbf{r})=-i\sum_je^{i\mathbf{\beta}\mathbf{R}_j}\sum_{lm}[a_l^m
\mathbf{N}_l^m(\mathbf{r}-\mathbf{R}_j)+b_l^m\mathbf{M}_l^m(\mathbf{r}-\mathbf{R}_j)].&
\end{eqnarray}
In series (\ref{out}) the first and second terms presents TE and TM spherical vector EM fields,
respectively.

In absence of an incident wave Linton {\it et al} \cite{Linton_Zal} derived the homogeneous
matrix equation for the amplitudes $a_l^m$ and $b_l^m$:
%-----------------------------------------------------------------------------------------------(3)
\begin{eqnarray}\label{matrixL0}
&Z_{TE,l}^{-1}a_l^m-\sum_{\nu}(a_{\nu}^m\mathcal{A}_{\nu l}^{mm}+b_{\nu}^m\mathcal{B}_{\nu l}^{mm})
=0,&\nonumber\\
&Z_{TM,l}^{-1}b_l^m-\sum_{\nu}(a_{\nu}^m\mathcal{B}_{\nu l}^{mm}+b_{\nu}^m\mathcal{A}_{\nu l}^{mm})
=0,&
\end{eqnarray}
where summation over $\nu$ begins with $max(1,m)$, and
the so-called Lorenz-Mie coefficients are given by
%--------------------------------------------------------------------------------------------------(4)
\begin{eqnarray}\label{Mie}
&Z_{TE,l}=\frac{j_l(kR)[rj_l(k_0r)]_{r=R}'-j_l(k_0R)[rj_l(kr)]_{r=R}'}
{h_l(k_0R)[rj_l(kr)]_{r=R}'-j_l(kR)[rh_l(k_0r)]_{r=R}'},&\nonumber\\
&Z_{TM,l}=\frac{\epsilon j_l(kR)[rj_l(k_0r)]_{r=R}'-j_l(k_0R)[rj_l(kr)]_{r=R}'}
{h_l(k_0R)[rj_l(kr)]_{r=R}'-\epsilon j_l(kR)[rh_l(k_0r)]_{r=R}'},&
\end{eqnarray}
where $k=\sqrt{\epsilon}k_0$ and  $\epsilon$ is the dielectric constant of the spheres,
%--------------------------------------------------------------------------------------------------(5)
\begin{equation}\label{A}
\mathcal{A}_{l\nu}^{mm}=4\pi(-1)^mi^{\nu-l}\sqrt{\frac{\nu(\nu+1)}{l(l+1)}}
\sum_{p=|l-\nu|;l+\nu+p=even}^{l+\nu}(-i)^pg_{l\nu p}\mathcal{G}(l,m;\nu,-m;p)s_p,
\end{equation}
%------------------------------------------------------------------------------------------------(6)
\begin{equation}\label{B}
\mathcal{B}_{l\nu}^{mm}=\frac{2\pi(-1)^m}{\sqrt{l(l+1)\nu(\nu+1)}}
\sum_{p=|l-\nu|+1;l+\nu+p =odd}^{l+\nu-1}i^{\nu-l-p}\sqrt{\frac{2p+1}{2p-1}}\mathcal{H}(l,m;\nu,-m;p)s_p.
\end{equation}
The coefficients
%--------------------------------------------------------------------------------------------------(7)
\begin{equation}\label{g}
    g_{l\nu p}=1+\frac{(l-\nu+p+1)(l+\nu-p)}{2\nu(2\nu+1)}-\frac{(\nu-l+p+1)(l+\nu+p+2)}
    {2(\nu+1)(2\nu+1)},
\end{equation}
%---------------------------------------------------------------------------------------------------(8)
\begin{equation}\label{G}
\mathcal{G}(l,m;\nu,\mu;p)=\frac{(-1)^{m+\mu}}{\sqrt{4\pi}}\sqrt{(2l+1)(2\nu+1)(2p+1)}
\left(\begin{array}{ccc} l & \nu & p \cr m & \mu &-m-\mu\end{array}\right)
\left(\begin{array}{ccc} l & \nu & p \cr 0 & 0 & 0\end{array}\right)
\end{equation}
are expressed in terms of Wigner 3-j symbols,
%-------------------------------------------------------------------------------------------------(9)
\begin{equation}\label{H}
\mathcal{H}(l,m;\nu,-m;p)=\sum_{s=-1}^1\mathcal{G}_s(l,m;\nu,-m;p)
\end{equation}
with
%---------------------------------------------------------------------------------------------------(10)
\begin{eqnarray}\label{GG}
&\mathcal{G}_0(l,m;\nu,-m;p)=-2m |p|\mathcal{G}(l,m;\nu,-m;p-1),&\nonumber\\
&\mathcal{G}_{\pm 1}(l,m;\nu,-m;p)=\mp\sqrt{(\nu\pm m)(\nu\mp m+1)p(p-1)}
\mathcal{G}(l,m;\nu,-m\pm 1;p-1),&
\end{eqnarray}
and
%---------------------------------------------------------------------------------------------------(11)
\begin{equation}\label{sigma}
s_p=\lambda_{p0}\sum_{j=1}^{\infty}h_p(k_0j)(e^{i\beta j}+(-1)^pe^{-i\beta j}),
\end{equation}
where $\lambda_{lm}$ is normalization factor given in Appendix B.

The next step is to account for an incident plane wave
which can be expanded over vector spherical harmonics \cite{Bruning,Stratton}
%-----------------------------------------------------------------------------------------------(12)
\begin{eqnarray}\label{ins}
&\mathbf{E}^{\sigma}(\mathbf{r})=\sum_{l=1}^{\infty}\sum_{-l}^l[q_{lm}^{\sigma}\mathbf{M}_l^m(\mathbf{r})+
p_{lm}^{\sigma}\mathbf{N}_l^m(\mathbf{r})],&\nonumber\\
&\mathbf{H}^{\sigma}(\mathbf{r})=-i\sum_{l=1}^{\infty}\sum_{-l}^l[p_{lm}^{\sigma}\mathbf{M}_l^m(\mathbf{r})+
q_{lm}^{\sigma}\mathbf{N}_l^m(\mathbf{r})].&
\end{eqnarray}
Here index $\sigma$ stands for plane TE and TM waves.
%--------------------------------------------------------------------------------------------------(13)
\begin{eqnarray}\label{pqTE}
&p_{lm}^{TE}=-F_{lm}\tau_{lm}(\alpha),~~q_{lm}^{TE}=F_{lm}\pi_{lm}(\alpha),&\nonumber\\
&p_{lm}^{TM}=-iF_{lm}\pi_{lm}(\alpha),~~q_{lm}^{TM}=iF_{lm}\tau_{lm}(\alpha),&
\end{eqnarray}
$k_x=-k_0\sin\alpha, k_y=k_0\cos\alpha$,
%--------------------------------------------------------------------------------------------------(14)
\begin{eqnarray}\label{FF}
&F_{lm}=(-1)^mi^l\sqrt{\frac{4\pi(2l+1)(l-m)!}{(l+m)!}},&\nonumber\\
&\tau_{lm}(\alpha)=\frac{m}{\sin\alpha}P_l^m(\cos\alpha),&\nonumber\\
&\pi_{lm}(\alpha)=-\frac{d}{d\alpha}P_l^m(\cos\alpha).&
\end{eqnarray}
For a particular case of normal incidence $k_z=0$ and $\alpha=-\pi/2$ we obtain
the following from Eqs. (\ref{FF})
%-------------------------------------------------------------------------------------------------(16)
\begin{equation}\label{tau}
\tau_{lm}=-mP_l^m(0), \pi_{lm}=-\frac{d}{d\alpha}P_l^m(0).
\end{equation}

The general equation for the amplitudes $a_l^m$ and $b_l^m$ which describe the scattering  by a linear array of spheres
takes the following form
%---------------------------------------------------------------------------------------------------(16)
\begin{eqnarray}\label{matrixL}
&Z_{TE,l}^{-1}a_l^m-\sum_{\nu}(a_{\nu}^m\mathcal{A}_{\nu l}^{mm}+b_{\nu}^m\mathcal{B}_{\nu l}^{mm})
=q_{lm}^{\sigma},&\nonumber\\
&Z_{TM,l}^{-1}b_l^m-\sum_{\nu}(a_{\nu}^m\mathcal{B}_{\nu l}^{mm}+b_{\nu}^m\mathcal{A}_{\nu l}^{mm})
=p_{lm}^{\sigma}.&
\end{eqnarray}
Here the left hand term formulates explicitly the matrix $\widehat{L}$  in Eq. (\ref{LS}) and the right
hand term corresponds to
the vector of incident wave $\mathbf{\Psi}_{inc}$ in the space of vector
spherical functions notified by two integers $l$ and $ m$ and polarization $\sigma$.

%%%%%%%%%%%%%%%%%%%%%%%%%%%%%%%%%%%%%%%%%%%%%%%%%%%%%%%%%%%%%%%%%%%%%%%%%%%%%%%%%%%%%%%%%
\section{The diffraction continua of vector cylindrical modes}
%%%%%%%%%%%%%%%%%%%%%%%%%%%%%%%%%%%%%%%%%%%%%%%%%%%%%%%%%%%%%%%%%%%%%%%%%%%%%%%%%%%%%%%%%%%%%%%
Thanks to the axial symmetry of the array we can exploit the vector cylindrical modes for
description of the diffraction continua which are doubly degenerate in TM and TE polarizations $\sigma$.
The modes can be expressed through the scalar function $\psi$ \cite{Stratton}
%--------------------------------------------------------------------------------------------------(17)
\begin{equation}\label{psi}
\psi_{m,n}(r,\phi,z)=H_m^{(1)}(\chi_n r)e^{im\phi+ik_{z,n} z}.
\end{equation}
Then for the TE modes we have
%---------------------------------------------------------------------------------------------------(18)
\begin{eqnarray}\label{cylTE}
    &E_z=0, ~~H_z=\psi_{m,n},&\nonumber\\
    &E_r=\frac{ik_0}{\chi_n^2}\frac{1}{r}\frac{\partial \psi_{m,n}}{\partial \phi},
    ~~H_r=\frac{ik_z}{\chi_n^2}\frac{\partial \psi_{m,n}}{\partial r},&\nonumber\\
&E_{\phi}=\frac{-ik_0}{\chi_n^2}\frac{\partial \psi_{m,n}}{\partial r},
    ~~H_{\phi}=\frac{ik_z}{\chi_n^2}\frac{1}{r}\frac{\partial \psi_{m,n}}{\partial \phi},&
\end{eqnarray}
and for the TM modes we have
%-------------------------------------------------------------------------------------------------(19)
\begin{eqnarray}\label{cylTM}
    &E_z=\psi_{m,n}, ~~H_z=0,&\nonumber\\
    &E_r=\frac{i k_z}{\chi_n^2}\frac{\partial \psi_{m,n}}{\partial r},
    ~~H_r=\frac{-ik_0}{\chi_n^2}\frac{1}{r}\frac{\partial \psi_{m,n}}{\partial \phi},&\nonumber\\
&E_{\phi}=\frac{ik_z}{\chi_n^2}\frac{1}{r}\frac{\partial \psi_{m,n}}{\partial \phi},
    ~~H_{\phi}=\frac{ik_0}{\chi_n^2}\frac{\partial \psi_{m,n}}{\partial r},&
\end{eqnarray}
where
%---------------------------------------------------------------------------------------------------(20)
\begin{equation}\label{chi}
    \chi_n^2=k_0^2-k_{z,n}^2
\end{equation}
and
%---------------------------------------------------------------------------------------------------(21)
\begin{equation}\label{kz}
    k_{z,n}=\beta+2\pi n, ~~n=0, \pm1, \pm2, \ldots.
\end{equation}
In what follows we consider the BSCs in the diffraction continua specified
by two quantum numbers $m$ and $n$ where the $m$ is the result of the axial symmetry and $n$ is the
result of translational symmetry of the infinite linear array of the dielectric spheres. Note that
each diffraction continuum is doubly degenerate relative to the polarization $\sigma$. As a result of the interplay between
the frequency $k_0$ and the wave number $k_{z,n}$ the continua can be open ($\chi$ is real) or closed ($\chi$ is imaginary).
%1 correction
The axial symmetry of the system substantially simplifies the consideration of BSCs since the azimuthal
behavior is specified by the integer $m$ only. That reduces the dimensionality of the system from
the 3D space of variables $r,\phi$ , and $z$ to the 2D space of $r$ and $z$.
%%%%%%%%%%%%%%%%%%%%%%%%%%%%%%%%%%%%%%%%%%%%%%%%%%%%%%%%%%%%%%%%%%%%%%%%%%%%%%%%%%%%%%%%%%%%%%%%
\section{Symmetry classification of BSCs}
%%%%%%%%%%%%%%%%%%%%%%%%%%%%%%%%%%%%%%%%%%%%%%%%%%%%%%%%%%%%%%%%%%%%%%%%%%%%%%%%%%%%%%%%%%%%%%%%%
In the previous section we presented the theory for the scattering of plane waves by a periodic array of dielectric spheres
based on the approach by Linton {\it et al} \cite{Linton_Zal}.
If there is no incident wave we have $\widehat{L}\mathbf{a}=0$
whose solutions are bound modes of the array. There might be two kinds of the
bound modes. The first type of modes  have wave number $\beta>k_0$ and describe guided waves along the array.
These solutions found by  Linton {\it et al} exist in some interval of the material parameters
of spheres, dielectric constant $\epsilon$ or radius $R$, and the Bloch wave number $\beta$
\cite{Linton_Zal}. The second type of bound modes with $\beta < k_0$
resides above the light cone (BSCs).
It is much more difficult to establish the existence of the second type of bound states
because a tuning of material parameters is required. However there might exist symmetry
protected BSCs which are robust with respect to the material parameters.
These BSCs have been already considered in the linear array of infinitely
long dielectric rods \cite{Shipman,Ndangali,Wei,Lopez,Bo Zhen,PRA,Song}.

The axial symmetry of the array implies that the matrices $\mathcal{A}$ and $\mathcal{B}$ split into
the irreducible representations of the azimuthal number $m$ which therefore classifies the BSCs.
Next, the discrete translational symmetry along the z-axis implies that the respective wave number
$\beta$ specifies the BSC. At last, additional optional symmetries arise due to the inversion
symmetry transformation $\widehat{K}f(x,y,z)=f(x,y,-z)$ for $\beta=0$ and $\pi$. It follows from
Eq. (\ref{sigma}) that $s_{2k+1}=0$, and respectively from Eqs. (\ref{A}) and (\ref{B}) we obtain
$\mathcal{A}_{\nu L}^{mm}=0$ if $l+\nu$ is odd, and $\mathcal{B}_{\nu L}^{mm}=0$ if $l+\nu$ is even.
Moreover for  arbitrary $\beta$: $\mathcal{B}_{\nu l}^{00}=0$
(see Appendix B). These relations establish the selection rules for
the amplitudes $a_l^m$ and $b_l^m$ which determine the allowed BSC modes listed in Table I.\\
\\ \ \\
\begin{center}
Table I. Classification of the BSCs\\
\begin{tabular}{|c|c|c|c|}
%\label{table 1}
  \hline\hline
$m$ &  $\beta$   & Type I of BSC& Type II of BSC \\
  % after \\: \hline or \cline{col1-col2} \cline{col3-col4} ...
\hline
  $\neq 0$ & 0 & $(a_{2k}^m, ~b_{2k+1}^m)$ & $(a_{2k+1}^m, ~b_{2k}^m)$ \\
  \hline
   0 &$\neq 0$ & $(a_l^0, ~0), E_z=0$ & $(0, ~b_l^0), H_z=0$ \\
   \hline
     0 & 0& $(a_{2k}^0, ~0), E_z=0$ & $(0, ~b_{2k}^0), H_z=0$ \\
\hline
     0 & 0& $(0, ~b_{2k+1}^0), H_z=0$ & $(a_{2k+1}^0, ~0), E_z=0$ \\
     \hline\hline
\end{tabular}
\end{center}

The cCartesian components of the vector spherical functions  transform under the inversion
of $z$ as follows
%---------------------------------------------------------------------------------------------------(22)
\begin{eqnarray}\label{MrNrcar}
&M_{l,x,y}^m(\pi-\theta)=-(-1)^{l-m}M_{l,x,y}^m(\theta), ~~
M_{l,z}^m(\pi-\theta)=(-1)^{l-m}M_{l,z}^m(\theta),&\nonumber\\
&N_{l,x,y}^m(\pi-\theta)=(-1)^{l-m}N_{l,x,y}^m(\theta),~~
N_{l,z}^m(\pi-\theta)=-(-1)^{l-m}N_{l,z}^m(\theta).&
\end{eqnarray}
For $\beta=0$ we have
%---------------------------------------------------------------------------------------------------(23)
\begin{eqnarray}\label{summ}
    &\sum_jM_{l,x,y}^m(\mathbf{r}-\mathbf{R}_j)=-(-1)^{l-m}
    \sum_jM_{l,x,y}^m(\widehat{K}\mathbf{r}-\mathbf{R}_j),&\nonumber\\
&\sum_jM_{l,z}^m(\mathbf{r}-\mathbf{R}_j)=(-1)^{l-m}
    \sum_jM_{l,z}^m(\widehat{K}\mathbf{r}-\mathbf{R}_j).&\nonumber\\
&\sum_jN_{l,x,y}^m(\mathbf{r}-\mathbf{R}_j)=-(-1)^{l-m}
    \sum_jN_{l,x,y}^m(\widehat{K}\mathbf{r}-\mathbf{R}_j),&\nonumber\\
&\sum_jN_{l,z}^m(\mathbf{r}-\mathbf{R}_j)=(-1)^{l-m}
    \sum_jN_{l,z}^m(\widehat{K}\mathbf{r}-\mathbf{R}_j).&
\end{eqnarray}
Then from these equations and Eqs. (\ref{out}) one can obtain
the following symmetric properties for the cartesian components of the EM fields
collected in Table II.\\ \ \\
\begin{center}
Table II. Symmetry properties of the eigenmodes with $\beta=0$.\\
\begin{tabular}{|c|c|}
%\label{table 1}
  \hline\hline
  % after \\: \hline or \cline{col1-col2} \cline{col3-col4} ...
  Type I& Type II\\
  \hline
  $E_{x,y}(-z)=(-1)^{m+1}E_{x,y}(z)$ & $E_{x,y}(-z)=(-1)^mE_{x,y}(z)$ \\
$E_z(-z)=(-1)^mE_z(z)$ & $E_z(-z)=(-1)^{m+1}E_z(z)$ \\
    $H_{x,y}(-z)=(-1)^mH_{x,y}(z)$ & $H_{x,y}(-z)=(-1)^{m+1}H_{x,y}(z)$ \\
$H_z(-z)=(-1)^{m+1}H_z(z)$ & $H_z(-z)=(-1)^mH_z(z)$ \\
  \hline\hline
\end{tabular}
\end{center}
Tables I and II are for the symmetry classification of the bound modes in the next sections.
%correction 2I will be
In particular, as it is seen from Table I for $m=0$ and  $\beta=0$ the type I of BSCs is the pure TE modes while
the type II is the pure TM modes. However when $m\neq 0$ the BSCs are given by superposition of
TE and TM polarized modes. Nevertheless each type I and type II of the BSCs presents a sort of
polarization because of their orthogonality to each other.
%%%%%%%%%%%%%%%%%%%%%%%%%%%%%%%%%%%%%%%%%%%%%%%%%%%%%%%%%%%%%%%%%%%%%%%%%%%%%%%%%%%%%%%%
\section{Symmetry protected BSCs}
%%%%%%%%%%%%%%%%%%%%%%%%%%%%%%%%%%%%%%%%%%%%%%%%%%%%%%%%%%%%%%%%%%%%%%%%%%%%%%%%%%%%%%%%%%%%
In this section we present numerical solutions of Eq. (\ref{matrixL0})
for the symmetry protected BSCs with $m=0$ and $\beta=0$ embedded into the
first diffraction continuum with $n=0$.
They constitute the majority of the BSCs in the array. The symmetry protected BSCs
are either pure TE spherical vector modes (type I in Table I) with $a_{2k}^0\neq 0$
and $ b_{k}^0=0$
or TM spherical vector modes (type II in Table I) with $a_k^0=0$ and $b_{2k}^0\neq 0$. We show
that the symmetry protected BSCs are symmetrically mismatched to the first open continuum.
\subsection{BSCs with $m=0$}
\begin{figure}[ht]
\includegraphics[scale=0.5,clip=]{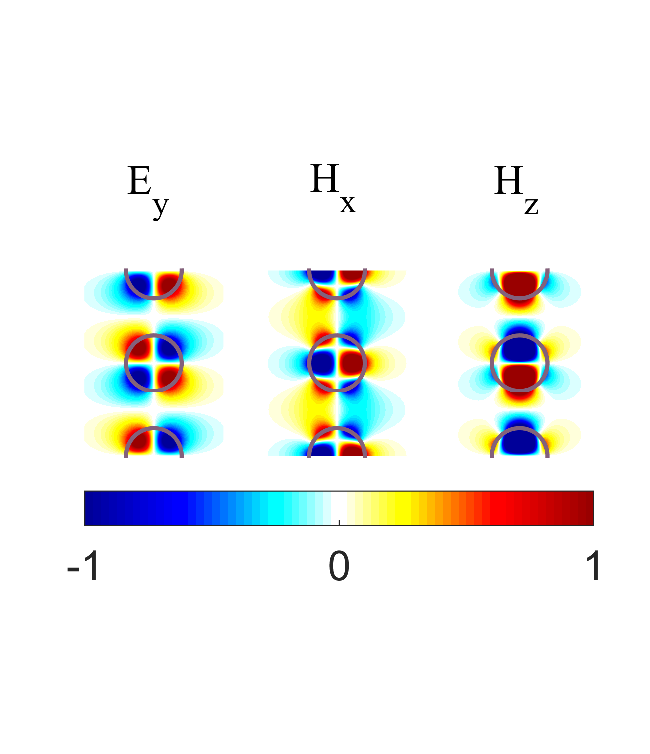}
\includegraphics[scale=0.35,clip=]{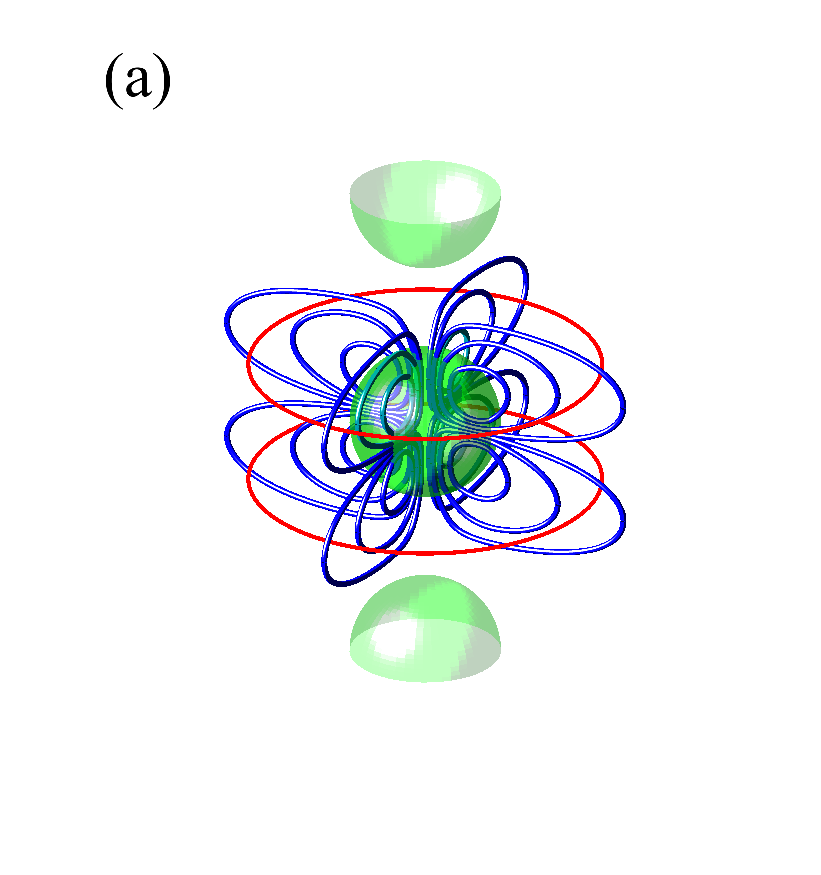}\\
\includegraphics[scale=0.5,clip=]{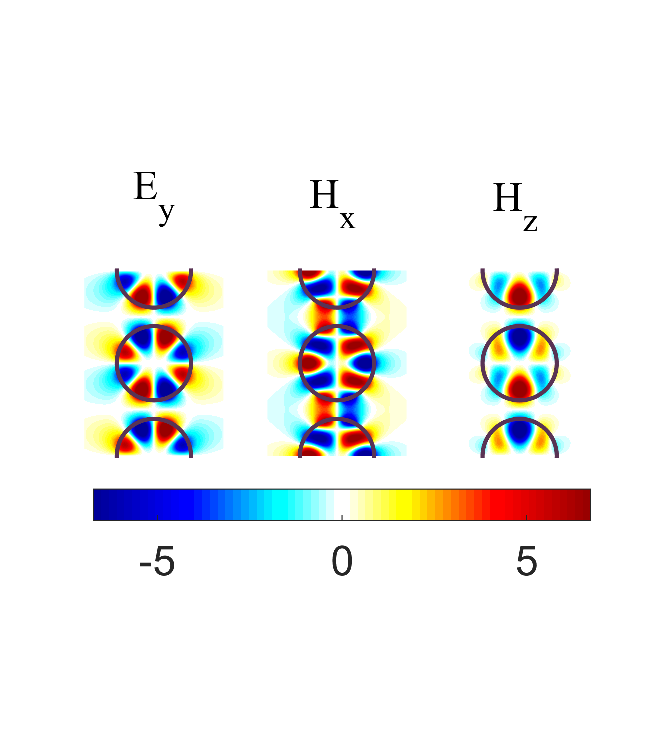}
\includegraphics[scale=0.35,clip=]{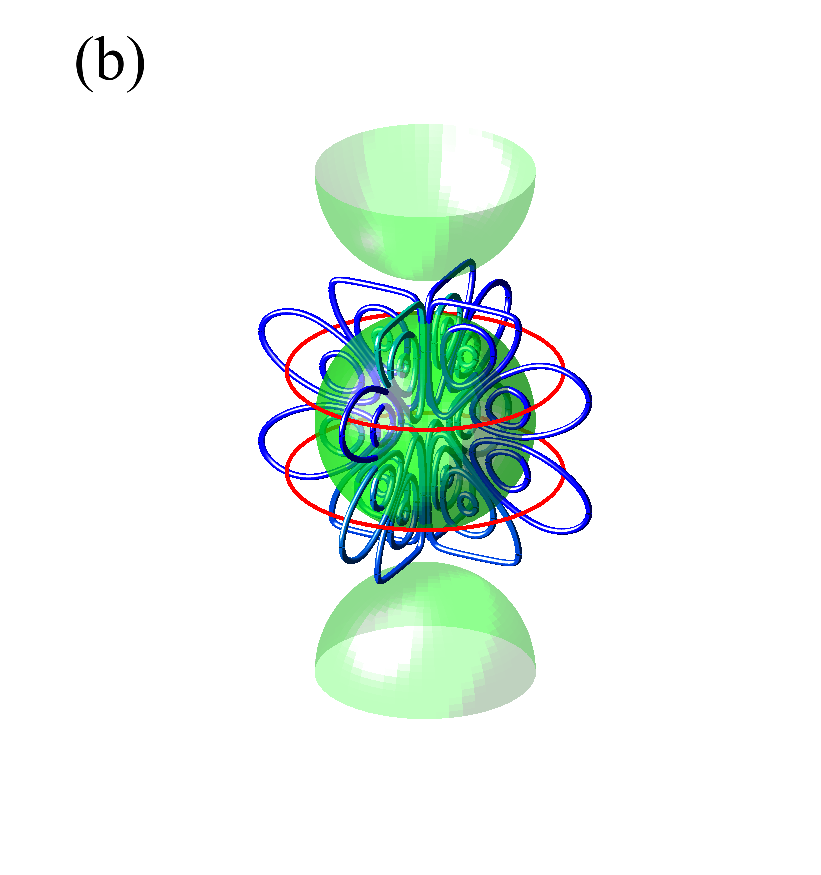}\\
\includegraphics[scale=0.5,clip=]{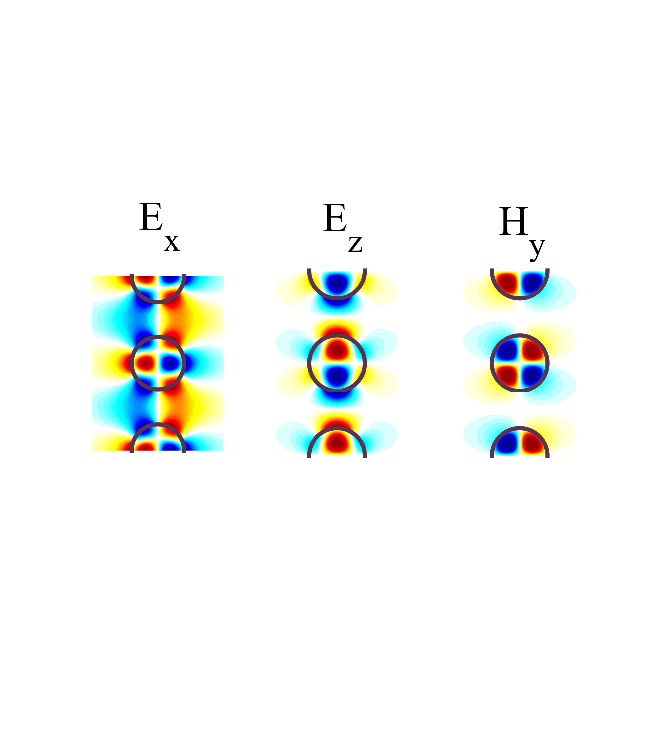}
\includegraphics[scale=0.4,clip=]{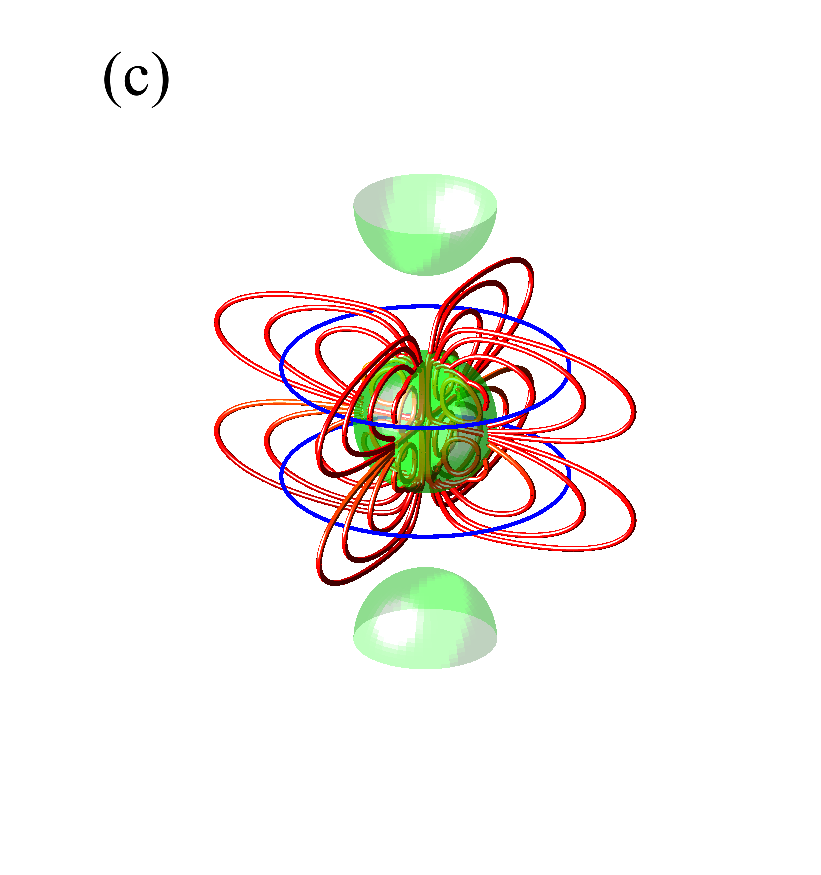}\\
\includegraphics[scale=0.3,clip=]{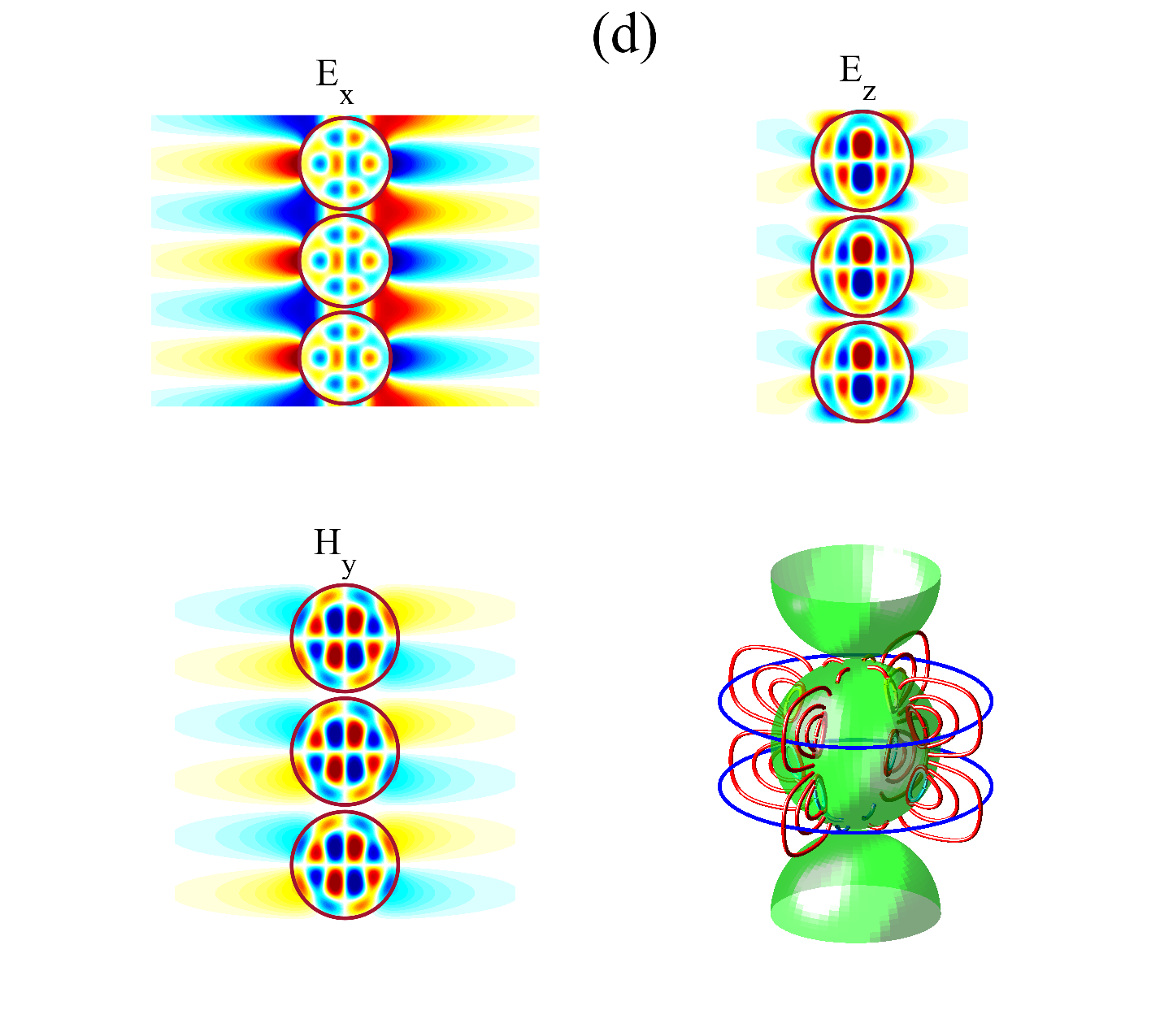}
\caption{Patterns of the symmetry protected  BSCs embedded into the TE and TM continua $m=0, n=0$.
TE BSCs (a) (\ref{BSC1}) and (b) (\ref{BSC2}) with electric force lines (red) parallel
to sphere surface. TM BSCs (c) (\ref{BSC16}) and (d) (\ref{BSC19})
with magnetic force lines (blue) parallel to sphere surface.
Left panels show the real parts of EM field components, right panels show the electric
force lines in red and magnetic force lines in blue.}
\label{fig2}
\end{figure}

Below we present numerical solutions for the TE BSCs with an accuracy of $10^{-4}$:
%---------------------------------------------------------------------------------------------------(24)
\begin{eqnarray}
\label{BSC1}
 k_0=4.24, ~~R=0.3, \epsilon=12,~~
a_l^0=
     \left(\begin{array}{c} 0 \cr  0.7563-0.6542i \cr 0 \cr
\end{array}\right),~~b_l^0=0, l\geq 1\\
   \label{BSC2}
   %-------------------------------------------------------------------------------------------------(25)
 k_0=5.0115, ~~R=0.4, \epsilon=12,~~
a_l^0=     \left(\begin{array}{c} 0  \cr  0.08576+0.1161i \cr 0 \cr
0.588+0.796i \cr 0 \cr -0.0002-0.0003i\end{array}\right),~~b_l^0=0, l\geq 1.
\end{eqnarray}
These TE BSCs embedded into the lowest $n=0$ diffraction continua of both polarizations
are shown in Figs.
\ref{fig2} (a) and (b). Hereinafter
we plot only the real parts of EM fields.
%-----------------------------------------------------------------------------------------------------Fig.2
There are also the TM BSCs:
%---------------------------------------------------------------------------------------------------(26)
\begin{eqnarray}
\label{BSC16}
k_0=4.7504, ~~R=0.3, \epsilon=15,~~a_l^0=0,~b_l^0=\left(\begin{array}{c} 0 \cr
 -0.6017+0.7988i\cr  0\cr  0.0004-0.0006i \end{array}\right), l\geq 1,\\
 \label{BSC19}
 %-----------------------------------------------------------------------------------------------------(27)
k_0=6.1522, ~~R=0.47, \epsilon=15,~~a_l^0=0,~b_0^1=
     \left(\begin{array}{c}  0 \cr  -0.8718+0.3926i\cr 0\cr
 0.267-0.1203i\cr  0\cr  -0.0013+0.0006i\end{array}\right), l\geq 1.
 \end{eqnarray}
Patterns of these TM BSCs are shown in  Figs. \ref{fig2}(c) and \ref{fig2}(d).
Due to the immediate  vicinity of the BSC (\ref{BSC19}) to the second
diffraction continuum the BSC shows a large scale of
localization around the spheres (see also the Sec. VIII below).

The symmetry protected TE and TM polarized BSCs have qualitatively similar field structure with respect to
$\mathbf{E}\leftrightarrow \mathbf{H}$
but are not degenerate because of different boundary conditions for $\mathbf{E}$ and $\mathbf{H}$
at the sphere surface.
The TE polarized BSC (\ref{BSC1}) and the TM polarized BSC (\ref{BSC16})
have the dominant contribution $a_2^0$  while the TE BSC
(\ref{BSC2}) and the TM BSC (\ref{BSC19}) have the noticeable contribution of $a_4^0$
which is reflected in complication
of the EM force lines shown in Figs. \ref{fig2}(b) and Fig. \ref{fig2}(d).
From Table II one can see why the eigenmodes
(\ref{BSC1})-(\ref{BSC19}) are protected by symmetry against decay into the diffraction continua $m=0, n=0$
with TE and TM polarizations.
From Eqs. (\ref{cylTE}) and (\ref{cylTM}) we obtain that the TE/TM continuum with $k_{z,0}=0 ~~(\beta=0)$ has the only
$H_z/E_z \neq 0$ independent of $z$. The TE BSC has $E_z=0$ and odd $H_z$ so that
this type of BSCs is symmetrically mismatched to both TE and TM continua.
Similarly, the TM BSC has odd $E_z$ and $H_z=0$ and is decoupled from both TE and
TM continua.

\begin{figure}[ht]
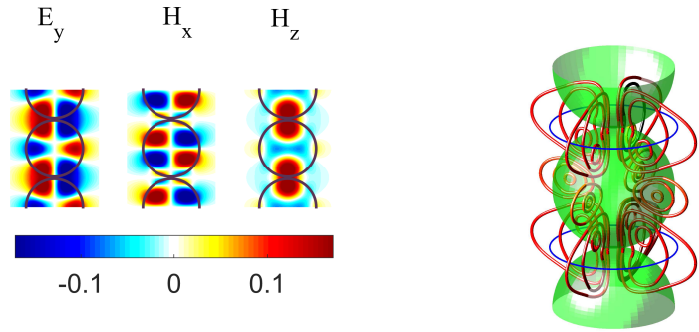

\includegraphics[scale=0.5,clip=]{BSC9.eps}
\includegraphics[scale=0.3,clip=]{EMBSC9.eps}
\caption{ Pattern of the TM BSC (\ref{BSC9}) embedded into the TE and TM diffraction continua
$m=0$ and $n=0$.}
\label{fig3}
\end{figure}

Besides the fully symmetry protected BSCs
from the third row in  Table I $(a_{2k}^0, 0)$ and $(0, b_{2k}^0)$, we found a
partially symmetry protected
TM BSC $(a_{2k+1}^0, 0)$ from the fourth row of Table I:
%----------------------------------------------------------------------------------------------------------------------------------------(28)
\begin{equation}
\label{BSC9}
 k_0=2.934, ~R=0.4805, ~\epsilon=15,~~
a_l^0=\left(\begin{array}{c} 0.6826+0.0332i \cr 0\cr  -0.7291-0.0354i \cr 0 \cr
-0.0008\end{array}\right),~~b_l^0=0, l\geq 1,
\end{equation}
however  the TE BSCs with $(0, b_{2k+1}^0)$ were not revealed in our computations.
The TM BSC (\ref{BSC9}) is symmetrically mismatched relative only to the $m=0$ and $n=0$ continuum with TM polarization.
Zero coupling of this BSC with the TE continuum can be achieved by tuning the radius of spheres.
Patterns of EM fields and EM force lines for this TM BSC are shown in  Fig. \ref{fig3}.
%---------------------------------------------------------------------------------------------------Fig.3

%%%%%%%%%%%%%%%%%%%%%%%%%%%%%%%%%%%%%%%%%%%%%%%%%%%%%%%%%%%%%%%%%%%%%%%%%%%%%%%%%%%%
\subsection{$\pm m$ degenerate BSCs with $\beta=0$}
\begin{figure}[httb]
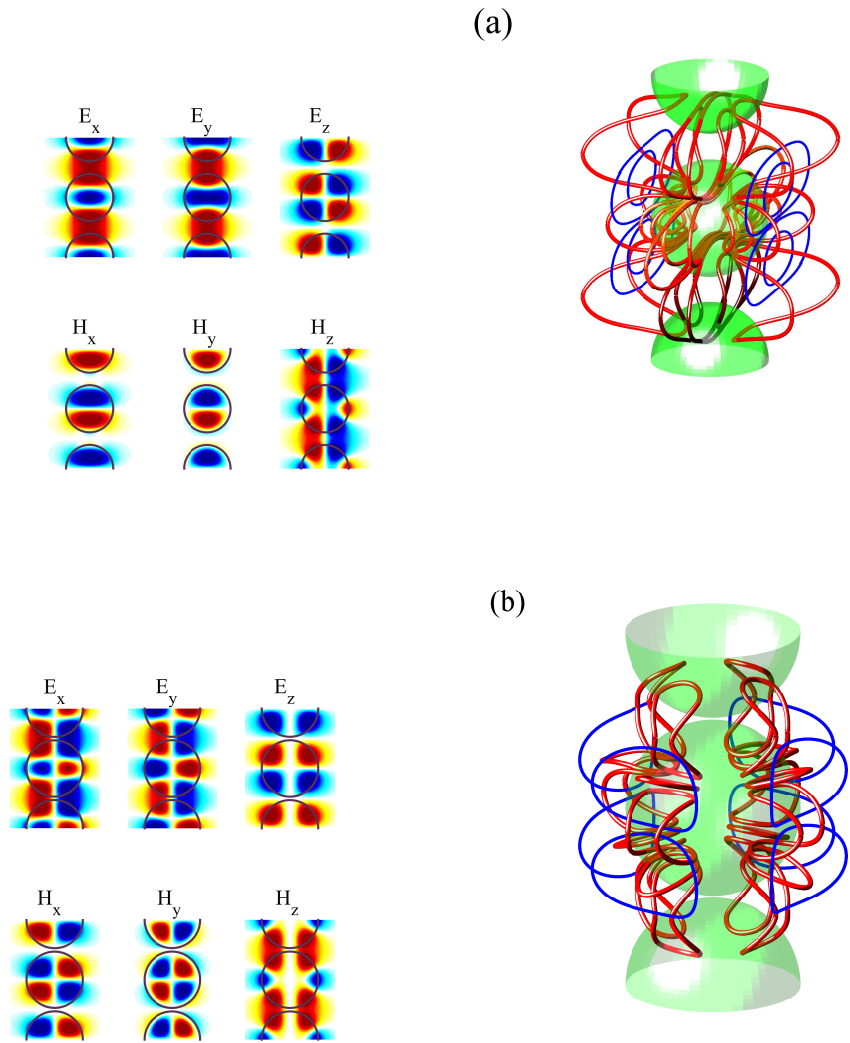

\includegraphics[scale=0.4,clip=]{BSC11.eps}
\includegraphics[scale=0.45,clip=]{EMBSC11.eps}\\
\includegraphics[scale=0.4,clip=]{BSC14.eps}
\includegraphics[scale=0.35,clip=]{EMBSC14.eps}
\caption{(Color online). BSC with $\beta=0$ embedded into the TE and TM continua $n=0$
and $m\neq 0$:
(a) the Ttype I BSC  (\ref{BSC11}) with $m=\pm 1$ and (b) the type II BSC (\ref{BSC14}) with $m=\pm 2$.}
\label{fig4}
\end{figure}

%%%%%%%%%%%%%%%%%%%%%%%%%%%%%%%%%%%%%%%%%%%%%%%%%%%%%%%%%%%%%%%%%%%%%%%%%%%%%%%%%%%%%
%Correction
For the $m=0$ case the BSC solutions can be described by
purely TE or TM modes in cylindrical coordinate as is shown in Figs. \ref{fig2} and
\ref{fig3}.
The case $m\neq 0$ is fundamentally different from the former case. Nevertheless
the above described mechanism for partially symmetry protected  BSCs with $m=0$ can be exploited
for even the case $m\neq 0$.
Obviously, the  system has the time reversal symmetry which implies that such BSCs are degenerated
over $\pm m$. Let us start with the type I BSC with $m=1$ which has the odd $E_z$ and the even $H_z$
according to Tables I and II. This BSC is symmetrically mismatched with
the TM diffraction continuum  $m=1$ and $n=0$ which is independent of $z$.
The coupling with the TE continuum can be canceled by tuning the radius.
The result of computation
of this partially symmetry protected type I BSC $(a_{2k}, b_{2k+1})$ is the following
%------------------------------------------------------------------------------------------------------------------------(29)
\begin{equation}
\label{BSC11}
 m=1, ~~    k_0=2.847, ~~R=0.3945, ~(a_l^1,~b_l^1)=
     \left(\begin{array}{cc} 0 & 0.6662+0.4273i\cr
     -0.33+0.5145i &0\cr
0 & -0.0048-0.0031i\cr 0 &0\cr \end{array}\right), ~l\geq 1
\end{equation}
and is shown in Fig. \ref{fig4} (a).
The type II BSC $(a_{2k+1}, b_{2k})$ with $m=2$ has  even $E_z$ and odd $H_z$.
It is symmetry protected against decay into the TE  continuum with $m=2$ and $n=0$
and coupling with the TM continuum is canceled by tuning  the radius with the following result
%---------------------------------------------------------------------------------------------------(30)
\begin{equation}
\label{BSC14} m=2, ~~k_0=3.086, ~~R=0.471, ~(a_l^2,~b_l^2)=
     \left(\begin{array}{cc} 0 & 0.6545+0.2013i \cr
 -0.2142+0.6964i & 0\cr 0 &-0.0057-0.0018i\cr
  0 & 0\cr 0& 0  \end{array}\right), l\geq 2.
\end{equation}
%-------------------------------------------------------------------------------------------------------Fig.4
All components of electric and magnetic fields are nonzero and localized around the array as
shown in Fig. \ref{fig4}. We show the EM field around only one sphere because the pattern is
periodically repeated along the z-axis. One can see that the value of the azimuthal
number $m$ is reflected
in the structure of force lines in the xy-plane while the number of the amplitudes  $a_l^m$ reflects in the structure
of lines along the z-axis. Figure \ref{fig4} clearly shows that the BSCs with $m\neq 0$ are neither TE
polarized nor TM polarized.
%%%%%%%%%%%%%%%%%%%%%%%%%%%%%%%%%%%%%%%%%%%%%%%%%%
\section{Robust Bloch BSCs with $\beta\neq 0$ and $m=0$}
%%%%%%%%%%%%%%%%%%%%%%%%%%%%%%%%%%%%%%%%%%%%%%%%%%%%%%%%%%%
Could the Bloch BSC occur at $\beta \neq 0$ in the continuum of free-space modes?
This question was first answered positively by Porter and Evans \cite{Porter} who
considered acoustic trapping in an array of rods of rectangular cross-section.
Marinica {\it et al} \cite{Shabanov} demonstrated the existence of the Bloch BSC with $\beta\neq 0$
in two parallel dielectric gratings and Ndangali and Shabanov \cite{Ndangali} in two
parallel arrays of dielectric rods.
In a single array of rods positioned on the surface of bulk 2d photonic crystal multiple BSCs
with $\beta\geq 0$ were considered by Hsu {\it et al} \cite{Wei}. The Bloch BSCs in a single
array of cylindrical dielectric rods in air were also reported in Ref. \cite{PRA}.
Such traveling wave Bloch BSCs with
the eigenfrequencies above the light cone are interesting because the array serves as a waveguide
although only for fixed $\beta$ (see summary of BSCs in Fig. \ref{fig8})
in contrast to the bound states below the light cone \cite{Linton_Zal}.

\begin{figure}[ht]
\includegraphics[scale=0.5,clip=]{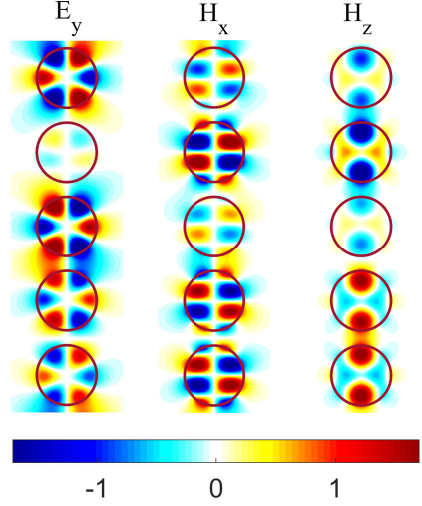}
\caption{(M field configurations of the TE Bloch BSC with $\beta=1.2074$ given
by Eq. (\ref{BSC15}) embedded into TE and TM continua $m=0, n=0$.}
\label{fig5}
\end{figure}

According to Table I  the Bloch BSCs with $\beta\neq 0$ $m=0$ have only the nonzero components $a_l^0$ or
$b_l^0$. Let us first consider type I BSCs with $b_l^0=0$ which have $E_z=0$ and, therefore, are decoupled
with the TM continuum but coupled with the TE $n=0$ and $m=0$ continuum.
We show numerically that this coupling can be canceled under variation
of $\beta$. The numerical results are collected in  Eq. (\ref{BSC15}) with the pattern of
EM fields shown in Fig. \ref{fig5}:
%The Bloch BSCs of the Type I or of the Type II with $m=0$ but
%$\beta\neq 0$ can not be classified as pure TE or TM modes respectively as different from the
%symmetry protected BSCs with $m=0, \beta=0$ discussed in Section V.
%----------------------------------------------------------------------------------------------------------------------(31)
\begin{equation}
%\label{BSC6}
 %k_0=3.571, ~~R=0.45, \epsilon=12,~\beta=0.8595, m=0, ~~(a_l^0,~b_l^1)=
  %   \left(\begin{array}{cc} -0.2164-0.1995i & 0 \cr
 %-0.33-0.358i & 0\cr 0.6046+0.5572i & 0\cr
  %-0.0078+0.0085i & 0\cr 0.0003+0.0003i& 0  \end{array}\right)\\
  \label{BSC15}
  k_0=3.6505, ~~R=0.4, \epsilon=15, ~\beta=1.2074, ~~(a_l^1,~b_l^1)=
     \left(\begin{array}{cc}
0.1053-0.0638i & 0 \cr
 0.1918+0.3161i & 0\cr 0.6046+0.5572i & 0\cr
  0.7873+0.4777i & 0\cr -0.0033-0.0054i& 0
     \end{array}\right), ~~l\geq 1.
\end{equation}
%-------------------------------------------------------------------------------------------------------Fig.5
Although this BSC occurs at the fixed value of $\beta$ there is no necessity
to tune the material parameters of the spheres and therefore the BSC can be referred to as
robust which is attractive from an experimental viewpoint. We managed to find only type I BSCs
for $\epsilon=15$ and none of
type II. Such a difference between the types  is related to different boundary conditions for electric and
 magnetic
fields at material interfaces.

%%%%%%%%%%%%%%%%%%%%%%%%%%%%%%%%%%%%%%%%%%%%%%%%%%%%%%%%%%%%%%%%%%%%%%%%%%%%%%%%%
\section{The bound states embedded into two and three diffraction continua}

\begin{figure}[ht]
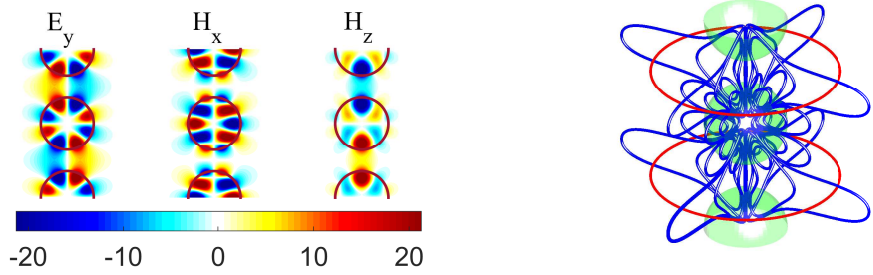

\includegraphics[scale=0.5,clip=]{BSC7.eps}
\includegraphics[scale=0.5,clip=]{EMBSC7new.eps}
\caption{Pattern of the TE BSC with $m=0$ and $\beta=\pi$ embedded into four TE and TM
continua with $m=0, n=0$ and $m=0, n=-1$ respectively given by Eq. (\ref{BSC7}).}
\label{fig6}
\end{figure}

According to  Sec. III the continua in the form of outgoing cylindrical waves
are specified by two numbers $m$ and $n$ which define $k_{z,n}$.
Above  we presented numerous BSCs embedded into the first diffraction continuum with $n=0$.
However there might be BSCs embedded into a few continua as it was shown for the case of grating
structures \cite{Ndangali,PRA}. Let us consider a TE BSC with $m=0$ and $\beta=\pi$ with
the nonzero components $a_{2k}^0\neq 0$,  $E_z=0$ and odd component $H_z$ according to Table I.
This BSC is coupled with the TE polarized radiation continua $m=0,n=0$ and $m=0,n=-1$ which
have $k_{z,0}=\pi$ and $k_{z,-1}=-\pi$ respectively. Because of degeneracy of
the continua we can form linear combinations with both even and odd $H_z$. Then, obviously, the
the BSC remains coupled with the odd $H_z$ continuum. This coupling can be canceled by variation of the sphere radius
to give the following result for the BSC amplitudes
\begin{equation}\label{BSC7}
        m=0, ~~\beta=\pi, ~~ k_0=5.0185, R=0.35456, \epsilon=15,
 a_l^0=
     \left(\begin{array}{cc} 0  \cr
 0.1552-0.0123i \cr 0 \cr
 -0.9847+0.0776i\cr 0 \end{array}\right), ~~l\geq 1.
        \end{equation}
The EM field and force lines are shown in Fig. \ref{fig6}.
%------------------------------------------------------------------------------------------------------Fig.6

\begin{figure}[ht]
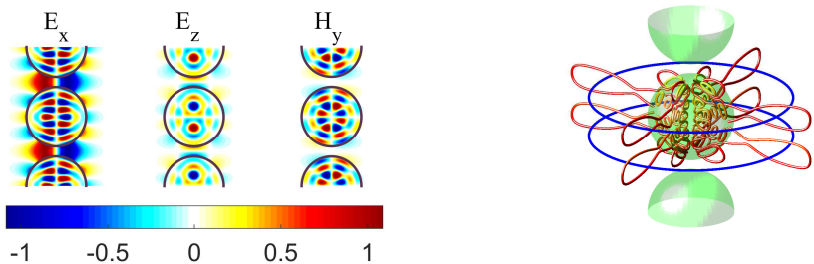

\includegraphics[width=6.5cm,clip=]{BSC10.eps}
\includegraphics[width=6.5cm,clip=]{EMBSC10.eps}
\caption{Pattern of the TM BSC with $m=0, \beta=0$ embedded into three continua
with $m=0, n=0$ and $m=0$ and $n=\pm 1$ given by Eqs. (\ref{BSC10}).}
\label{fig7}
\end{figure}

\begin{figure}[ht]
\includegraphics[scale=1,clip=]{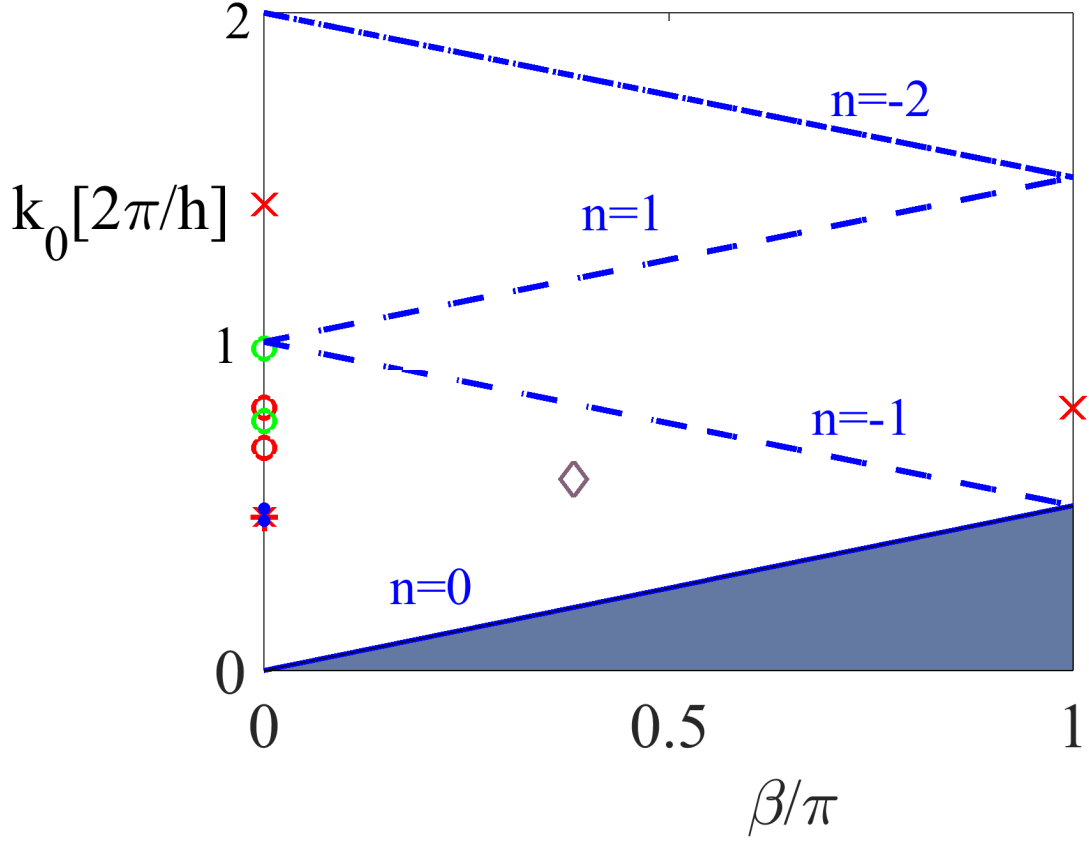}
\caption{(Color online). BSC frequencies and Bloch vector $\beta$ relative to
the light line $k_0=\beta$. Dash and dash-dot lines show thresholds where the next continua $n=\pm 1$ and
$n=-2$ are opened. Fully symmetry protected BSCs (\ref{BSC1}) and (\ref{BSC16}) are marked by open circles
(TE BSCs are in red color) and TM BSCs are in green color), TM BSC (\ref{BSC9}) is marked by star, two degenerate BSCs (\ref{BSC11}) and
(\ref{BSC14}) are marked by points,
Bloch BSC (\ref{BSC15}) with $\beta\neq 0$ is marked by a rhombus, and BSCs (\ref{BSC7}) and (\ref{BSC10}) embedded into two and three
continua are marked by crosses.}
\label{fig8}
\end{figure}

We also found a type II BSC $(0, b_{2k})$ with $m=0$ and $\beta=0$ with odd $E_z$ and $H_z=0$ embedded into three continua with
$n=0$ and $n=\pm 1$ shown in Fig. \ref{fig7}. As is shown above this BSC is completely decoupled from
the TE and TM radiation
%---------------------------------------------------------------------------------------------------------------Fig.7
continua with $n=0$ due to the symmetry. As for the other continua with $n=\pm 1$
the BSC  is decoupled from the TE continua. Similar to the previous case the degenerate TM continua
have $k_{z,\pm 1}=\pm 2\pi$ and   can be superposed into the continua
with either even or odd $E_z$. Thus, the Type II BSC is decoupled with the continuum with even $E_z$.
By  variation of the sphere radius we achieved zero coupling with the continuum with
odd $ E_z$ with the following solution:
\begin{equation}\label{BSC10}
        m=0, ~~\beta=0, ~~ k_0=8.9129, R=0.4274, \epsilon=15, ~~
b_l^0= \left(\begin{array}{cc}  0 \cr
 -0.2273-0.1508i\cr 0 \cr
 -0.8015-0.532i\cr  0\cr  0.0082+0.0055i\end{array}\right), ~~l\geq 1.
 \end{equation}
We collected the BSC frequencies $k_0$ and Bloch vectors $\beta$ in Fig. \ref{fig8}.
%---------------------------------------------------------------------------------------------------------Fig.8
%%%%%%%%%%%%%%%%%%%%%%%%%%%%%%%%%%%%%%%%%%%%%%%%%%%%%%%%%%%%%%%%%%%%%%%%%%
\section{Emergence of the BSC in scattering}
%%%%%%%%%%%%%%%%%%%%%%%%%%%%%%%%%%%%%%%%%%%%%%%%%%%%%%%%%%%%%%%%%%%%%%%%%%%%
Scattering of plane waves  by periodic 2D arrays of
dielectric spheres was first considered in the pioneering papers by Ohtaka
{\it et al} \cite{OhtakaJPC,Inoue,Miyazaki} (see also Ref. \cite{Modinos}).
Scattering by aggregates of spheres was considered in the framework of multi-sphere Mie scattering  \cite{FullerI,Bruning,Xu},
nevertheless to  our knowledge the scattering by a 1D infinite array of dielectric spheres has not been considered so far.
In this section we present the results of numerical computations for
differential and total cross-sections of the infinite array with the focus on
resonant traces of the BSCs similar to the scattering by an array
of dielectric rods \cite{Wei,PRA,Bo Zhen}. In what follows we restrict
ourselves to the BSCs which are standing localized solutions with
$\beta=0$. The general theory of scattering in terms $(a_l^m, b_l^m)$
is formulated in the form of  Eq. (\ref{matrixL}) which allows to find the amplitudes.
After the amplitudes are found from Eq. (\ref{matrixL}) one can expand
EM fields (\ref{out}) over vector cylindrical modes to calculate the cross-sections.

While the BSCs are given by the homogeneous part of Eq. (\ref{LS}) with
$\mathbf{\Psi}_{inc}=0$, the scattering fields
are given by the solution of inhomogeneous Eq. (\ref{matrixL0}) with
an incident plane wave at the right hand part. As it follows from
Eqs.(\ref{chi}) and (\ref{kz})  only one diffraction channel $n=0$
is open for low frequencies $k_0$ where the majority of the BSCs occur.
Although the BSCs can not be probed directly by an incident wave
they  are seen as collapses of the Fano resonance when the BSC point is approached in the parametric space.
That phenomenon  was observed
for the scattering of EM waves by  arrays of rods
\cite{Shipman,Ndangali,Wei,robust,PRA,Zhang,Bo Zhen,Song} and layered sphere \cite{Alu}. In this
section we report a similar Fano resonance collapse
in the differential and total cross-sections vs frequency when the wave number $k_z$ tends to zero or the radius of the spheres
approaches the BSC radius. The Fano resonance for the
present system can be interpreted as an interference of the optical paths through
and between the spheres. We restrict ourselves to the BSC effects on the cross-section for
the fully symmetry protected BSCs and the BSCs degenerate over $m=\pm 2$.

Let us consider an incident plane wave with the wave vector in the $x,z$ plane and polarizations:
(a) TE  polarized with the electric field along the y-axis and (b) TM polarized with the magnetic field
along the y-axis.
%-------------------------------------------------------------------------------------------------------Fig.9
\begin{figure}[ht]
\includegraphics[scale=0.5,clip=]{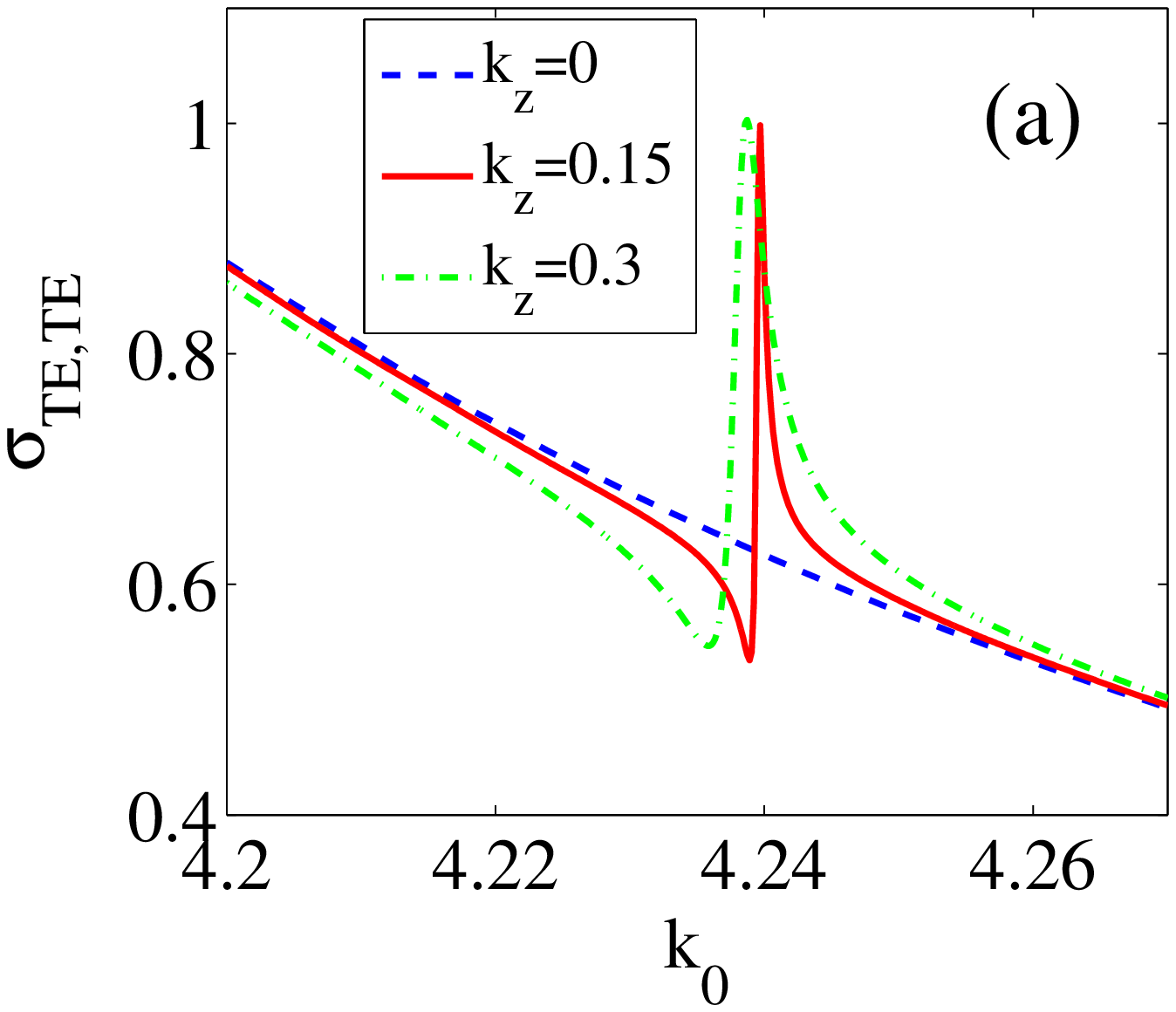}
\includegraphics[scale=0.5,clip=]{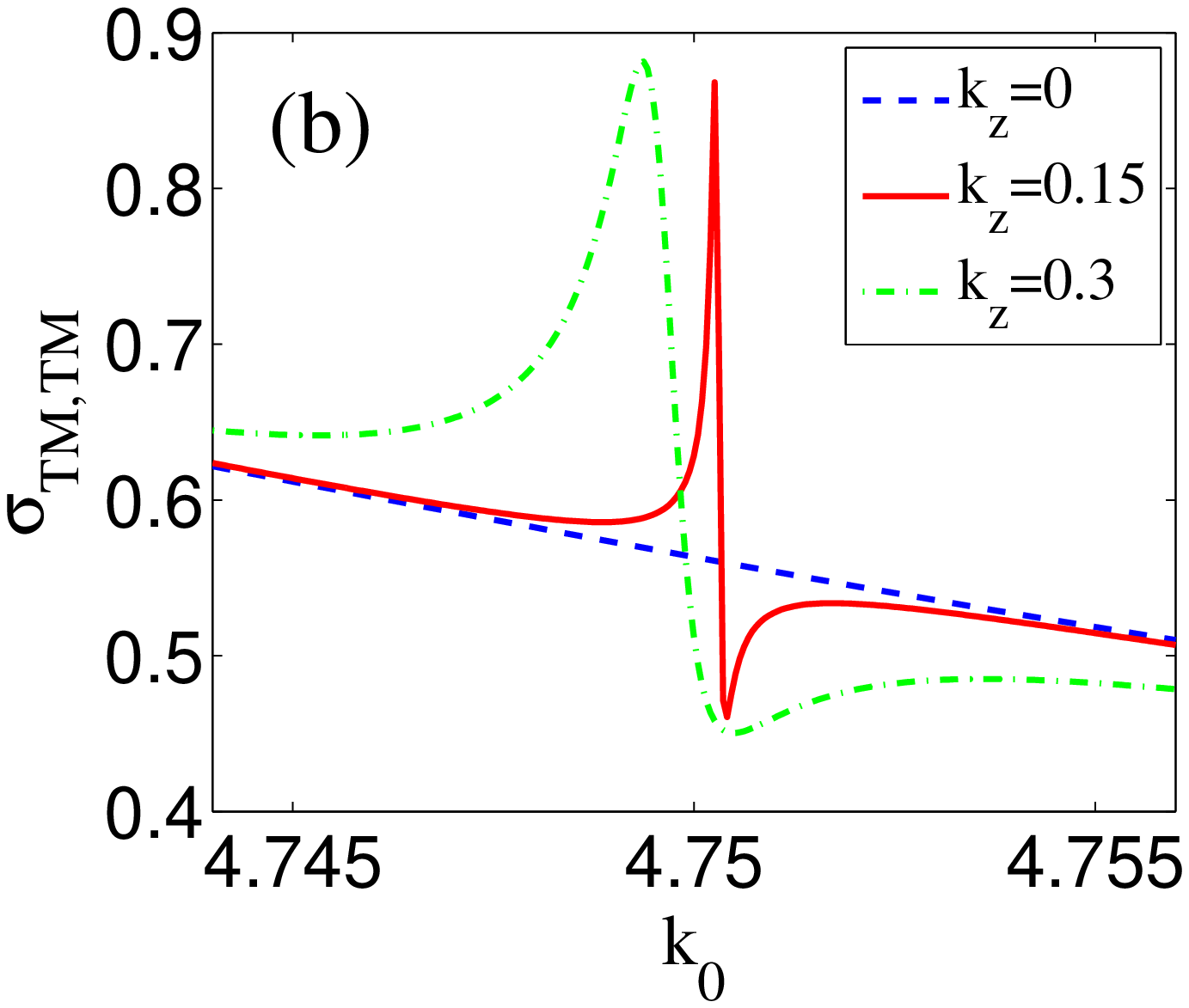}
\caption{Total cross-section for scattering of the
plane wave incident by the angle $\phi$ onto the array. (a) Scattering of the TE plane wave is strongly
affected by the presence of the symmetry protected type I BSC (\ref{BSC1}) with the
eigenfrequency $k_0=4.24$ for $R=0.3$ and $ \epsilon=12$. (b) Scattering
of the TM plane wave is strongly affected by the presence of the symmetry
protected type II BSC (\ref{BSC16}) with the eigenfrequency
$k_0=4.7504$ for $R=0.3$ and $ \epsilon=15$.} \label{fig9}
\end{figure}
For $m=0$ and $k_z\neq 0$ Eqs. (\ref{pqTE}) and (\ref{FF}) give that $p_{l0}^{TE}=0$ and $q_{l0}^{TM}=0$.
Then taking into account that $\mathcal{B}_{\nu l}^{00}=0$
(see Appendix B) we have the following from Eqs. (\ref{matrixL}) for the TE incident plane wave
%---------------------------------------------------------------------------------------------------(34)
\begin{eqnarray}\label{matrixLnor}
&Z_{TE,l}^{-1}a_l^0-\sum_{\nu}a_{\nu}^0\mathcal{A}_{\nu l}^{00}=q_{l0}^{TE},&\nonumber\\
&Z_{TM,l}^{-1}b_l^0-\sum_{\nu}b_{\nu}^0\mathcal{A}_{\nu l}^{00}=0.&
\end{eqnarray}
%--------------------------------------------------------------------------------------------------------Fig.10
\begin{figure}[ht]
\includegraphics[scale=0.5,clip=]{fig10a.eps}
\includegraphics[scale=0.5,clip=]{fig10b.eps}
\includegraphics[scale=0.5,clip=]{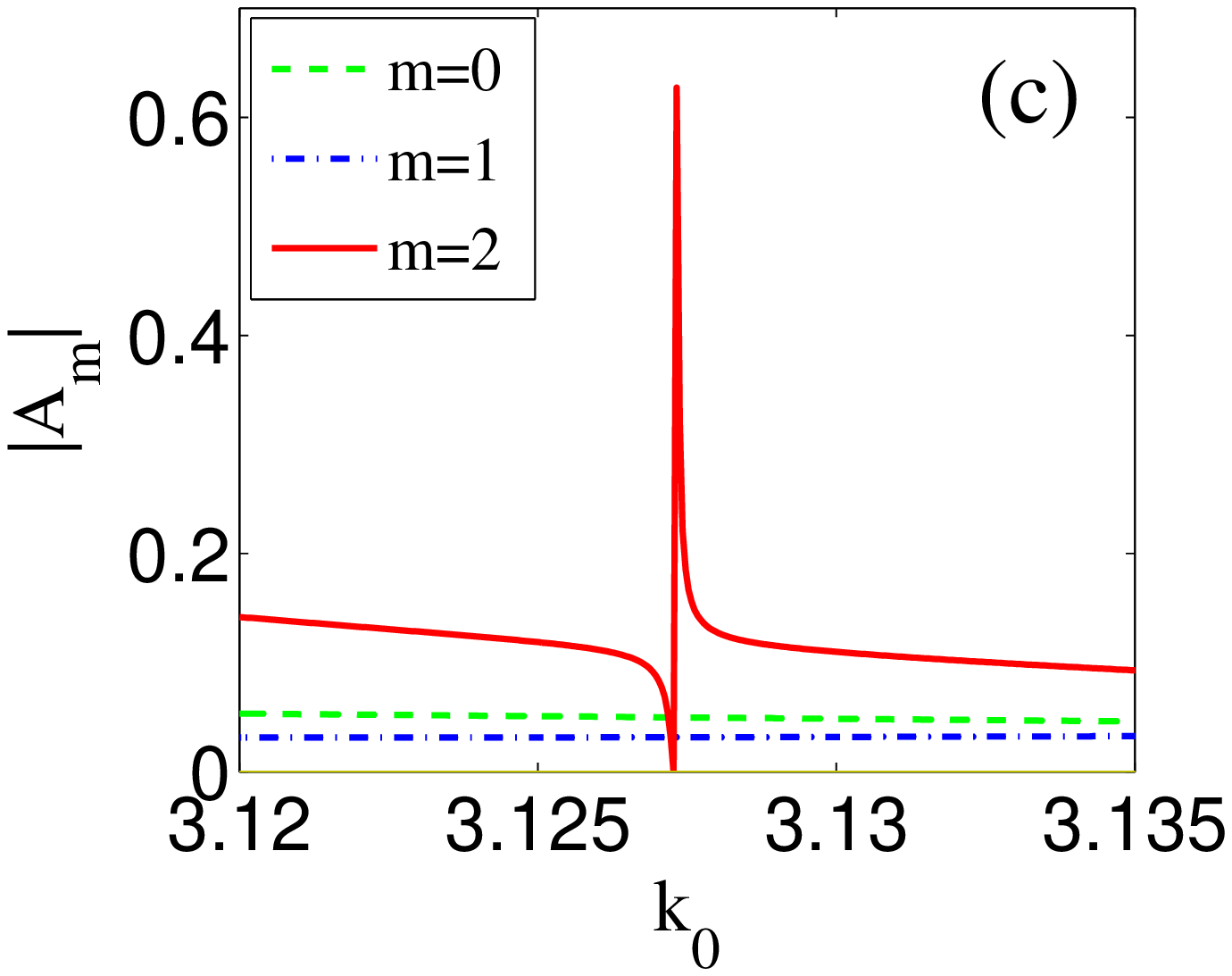}
\includegraphics[scale=0.5,clip=]{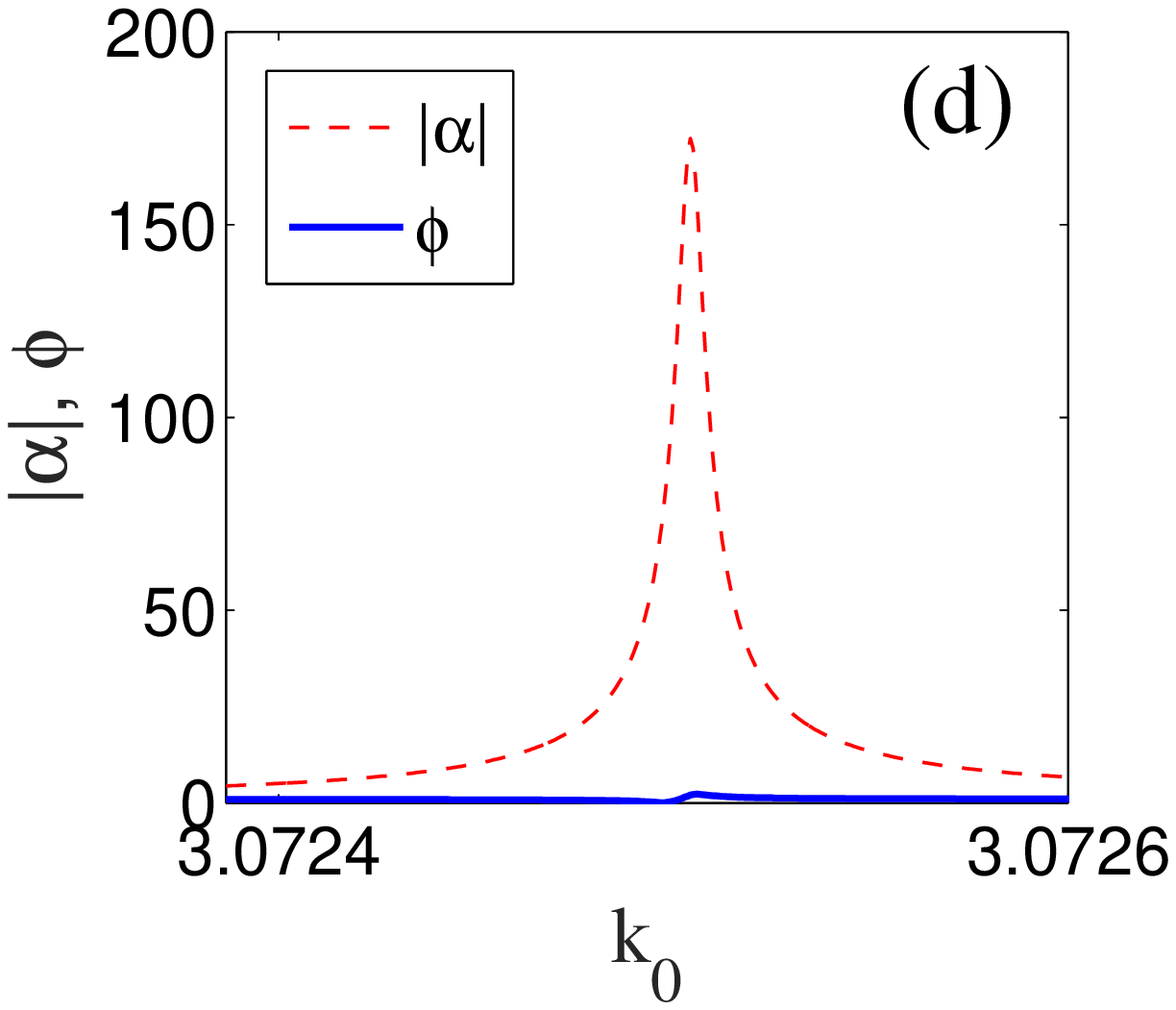}
\caption{The effect of the BSC (\ref{BSC14}) with
$m=2, k_0=3.086$ and $R=0.471$ in: (a)
differential cross-section vs frequency and the azimuthal angle,
(b) total cross-sections for different radii of the spheres
close to the BSC radius (\ref{BSC11}) for plane
wave illuminating the array normally. (c) Frequency behavior of the
amplitudes $A_m$  in the expansion (\ref{Am}). (d)
Harvesting capability of the quasi BSC at  $R=0.473$. The dashed red line shows the contribution of the BSC
into the scattering function; the blue solid line shows the background $\phi$.} \label{fig10}
\end{figure}
We do not present here the sectors of wave scattering with $m\neq 0$ since the type I BSC belongs to the sector $m=0$.
The second equation gives $b_l^0=0$, so that the plane wave with TE polarization after the scattering is given by $a_l^0$ only.
Then the type I BSCs are quasi BSCs weakly coupled with the TE continuum for small $k_z$.  That results in a sharp resonant contribution in the
cross-section $\sigma_{TE,TE}$ as shown in Fig. \ref{fig9} (a).
The cross-sections $\sigma_{TE,TM}, \sigma_{TM,TM}$ and $\sigma_{TM,TE}$
have no features related to these BSCs and are not shown in Fig. \ref{fig9} (a).
If the plane wave is incident onto the array normally
$\alpha=-\pi/2$ ($k_z=0$) we have a fully invisible type I BSC that is shown by dash line in Fig. \ref{fig9} (a).
 Alternatively, the  symmetry protected type II BSCs with the only nonzero
 $b_k$ can be observed via the cross-section $\sigma_{TM,TM}$ as shown
 in Fig. \ref{fig9} (b). Thus, although the BSCs have no effect for the normal incidence
they are detected by the collapse of Fano resonances in total cross-sections for $k_z\rightarrow 0$.

Next, consider the effect of the BSCs with  $m=2$ given by Eq. (\ref{BSC14}) on the cross-section.
We begin with the TE plane waves incident onto the array
normally ($k_z=0$). Then we have from Eqs. (\ref{pqTE})-(\ref{tau}) that $p_{l2}^{TE}=0, q_{l2}^{TE}\neq 0$
for odd $l$,
and $p_{l2}^{TE}\neq 0, q_{l2}^{TE}=0$ for even $l$. Therefore as Eq. (\ref{matrixL}) shows
there are only type II solutions for scattered waves with the amplitudes $(a_{2k+1}, b_{2k})$.
Table I shows that they  belong to the same type of  BSCs with $m=2$.
Therefore in the vicinity of $R_{BSC}=0.471$ this BSC is coupled with the TE continuum and
gives the resonant contribution in the cross-section $\sigma_{TE,TE}$  that
is demonstrated in Figs. \ref{fig10} (a) and (b).
As for the scattering of the TM plane waves there are no resonant features as shown in Fig. \ref{fig10} (b) by dash line.
One can see in Fig. \ref{fig10} (c) bright features of the differential
cross-sections near the eigenfrequency of the BSC caused by the resonant contribution of
the amplitude $A_2$ at the azimuthal angles $\phi=0, \pm 90^o, 180^o$:
%---------------------------------------------------------------------------------------------------------------------(35)
\begin{equation}\label{Am}
    \frac{d\sigma}{d\phi}=\sigma_0\vert\sum_m A_mcos(m\phi)\vert^2.
\end{equation}

It is clear that for the sphere radius close to  $R_{BSC}=0.471$ the BSC solution dominates in the near field zone.
The solution can be presented as
%--------------------------------------------------------------------------------------------------------------------------(36)
\begin{equation}\label{near}
    \mathbf{\Psi}=\alpha\mathbf{\Psi}_{BSC}+\mathbf{\Phi}
\end{equation}
where $\alpha$ has a resonant behavior over frequency $k_0$ with the resonant width
$\gamma\sim|R-R_{BSC}|$. Analytical expression for the resonant width can be  derived following Refs.
\cite{ring,SBR}. Thus we have slowly decaying quasi BSC modes above the light cone
similar to those considered in Ref. \cite{Ochiai}. That effect is important for concentration
of light by touching spheres \cite{Pendry,Zhang} notified as the harvesting capability of the system.
Fig. \ref{fig10} (d) illustrates the harvesting capability of the array of spheres in the vicinity of the
BSC (\ref{BSC14}). Solid blue line shows the contribution of the background $\phi=||\mathbf{\Phi}||$
where $||\cdots||$ is the norm of vector $\mathbf{\Phi}$.

\section{Summary}
Recently the BSCs above the light cone were shown to exist
in various systems of 1D arrays of dielectric rods and holes in a
dielectric slab  \cite{Shipman,Ndangali,Shabanov,Wei,Bo Zhen,Yang,PRA,Hu&Lu}.
Similar acoustic BSCs called embedded trapped Rayleigh-Bloch surface waves were obtained in a
system of material rods \cite{Porter0,Porter1,Porter,Linton,Colquitt}.
One could ask why BSCs occur in periodic dielectric structures (gratings) but not in homogeneous structures like
a slab or a rod which can support guided EM modes below the light cone only.
Let us begin with the simplest textbook system of a dielectric slab  infinitely long in the $x,y$ plane with
the dielectric constant $\epsilon>1$. The Maxwell equations  can be solved by separation of variables for scalar function
$\psi(x,y,z)=e^{ik_xx+ik_yy}\psi(z)$ to result in bound states below the light cone
$k_0^2=k_x^2+k_y^2$ \cite{Jackson} while all solutions above the light cone are leaky \cite{Hu1}.
The situation can be cardinally changed by replacing the continual translational symmetry by
the discrete symmetry $\epsilon(x,y,z)=\epsilon(x+ph,y,z)$ where $p=0, \pm 1, \pm 2,\ldots$, and $h$ is
the period of the structure.
Then the radiation continua of plane waves $e^{ik_{x,n}x+ik_yy+ik_zz}$ are quantized
$k_{x,n}=\beta+2\pi n/h, ~~n=0, \pm 1, \pm 2,\ldots$ with the frequency $k_0^2=k_{x,n}^2+k_y^2+k_z^2$.
Here $\beta$ is the Bloch wave vector along the x-axis, and the integer $n$ refers to the
diffraction continua \cite{Ndangali}. The physical interpretation of this statement is related to the
slab with the discrete translational symmetry being considered as a 1D diffraction lattice in the x-direction.
Let us take for simplicity $\beta=0$ and $k_y=0$. Assume
there is a bound solution with the eigenfrequency $k_{0,BSC}>0$ which is coupled with all diffraction
continua enumerated by $n$. Let $k_{0, BSC}<2\pi/h$, i.e.,
the BSC resides in the first diffraction continua but below the others.
Because of the symmetry or by variation of the material parameters of the modulated slab
we can achieve that the coupling of the solution with
the first diffraction continuum equals zero \cite{Wei,Bo Zhen,Yang,PRA,Hu&Lu}.
However the solution is coupled with evanescent continua $n=1,2, \ldots$
giving rise to exponential decay of the bound solution over the $z$-axis.
The length of localization is given by $L\sim\frac{1}{\sqrt{4\pi^2/h^2-{k_{0, BSC}^2}}}$.
Therefore the evanescent diffraction continua
play a principal role in the space configuration of the BSCs. Moreover, one can see from Fig. \ref{fig8} that
in the limit $h\rightarrow \infty$ the BSCs with frequency
$k_{0 BSC}\rightarrow 0$  leave no room for the BSCs with $k_{0,BSC}>0$.

In the present paper we choose another strategy to quantize the radiation continuum. We
replace the rod with continual translational symmetry by a periodic array of
dielectric spheres.  Because of the axial symmetry of the array aligned along the z-axis the quantized
continua are specified by two integers, $m$ and $n$.
The first integer is the azimuthal quantum number and the second number defines discrete directions of
outgoing cylindrical waves (\ref{psi}) given by the wave vector $k_{z,n}=\beta+2\pi n/h$ in each sector $m$
where $\beta$ is the Bloch vector along the array. The bottoms of the particular  continua with $m=0$ and $n=0, \pm 1$ and $n=-2$ are shown
in Fig. \ref{fig8}. By arguments similar to those presented above for the grated slab we obtain that
the BSC with $\beta=0$ embedded into the first radiation continuum $m=0, n=0$ is localized around the array
with the radius of localization given by $\frac{1}{\sqrt{4\pi^2/h^2-{k_{0, BSC}^2}}}$ .

The symmetry of the system is also important for classifications of the BSCs which are labeled by the azimuthal
number of the continuum $m$ of cylindrical vectorial waves and the  Bloch
wave vector $\beta$.

(1) The symmetry protected BSCs constitute the vast majority of BSCs
which are symmetrically mismatched with the first diffraction continuum with $m=0$ and $n=0$ of both
polarizations. The EM field configurations of such BSCs  presented in Fig.
\ref{fig2} show hybridizations of a few orbital numbers $l=2, 4, 6, \ldots$ which specify the BSCs as
multipoles of high order. Therefore the BSC solutions can not be obtained by the use of the dipole approximation
\cite{Draine,Savelev}.
The most remarkable property from an experimental viewpoint is the
robustness of the BSCs relative to the choice of the material parameters of the dielectric spheres.
We present in Fig. \ref{fig3} an example of the BSC which is symmetry protected
relative to the TM diffraction continuum but has a zero coupling to the TE continuum obtained through variation of
the sphere radius.

(2) By tuning of the radius of the spheres we found BSCs in the next sectors of continua with $m\neq 0$.
These BSCs  shown in Fig. \ref{fig4} are remarkable by degeneracy over the sign of the azimuthal number.
Each BSC with $\pm m$ has opposite the Poynting vector.

(3) We demonstrated  that the BSC can be accessed not only by variation of the material parameters
but also by variation of Bloch wave vector $\beta$ along the array axis. Patterns of the Bloch BSCs are
presented in Fig. \ref{fig3}.

(4) We found the
trapped EM modes embedded into two diffraction continua with $n=0$ and $n=1$ (Fig. \ref{fig6}) and
even three continua with $n=0$ and $n=\pm 1$ (Fig. \ref{fig7}).

The symmetry properties of the BSC play a very important role since it is difficult to provide a zero coupling
even with the lowest continua with $n=0$ because of the degeneracy in polarization.
Nevertheless the symmetry allows one to decouple
the BSC at least with some particular continua.

The advantage of dielectric structures is a high quality factor and a wide range of
BSC wavelengths from microns (photonics) to centimeters (microwave) as dependent on the choice of the radius
of the spheres. Although the BSCs exist only in selected points in the parametric space
there is a nearest vicinity  of the BSC point  where the BSC predominantly contributes into the cross-section
and the EM field in the near field zone as seen from Figs. \ref{fig9} and \ref{fig10}.
That leads to extremely efficient light harvesting capabilities \cite{Pendry}. The far zone EM fields
can also show abundant features related to the BSCs. In particular Fig. \ref{fig8} (a) demonstrates
the effect of antenna when the BSC with azimuthal number $m=2$ converts the EM energy
into the perpendicular directions.
%In practice the array of spheres can be fabricated of finite number $N$ of spheres that implies defects
%at the ends of surface. That gives rise to quasi BSCs with the finite life time which however is expected
%to be of order $N$. Since far from the ends of the array the quasi BSC is localized it will radiate light at the
%ends.

{\bf Acknowledgments}:

 The work was supported by the Russian Science Foundation through grant 14-12-00266.
 We acknowledge discussions with  D.N. Maksimov.

\section{Appendix A}
The solution of the Maxwell equations inside and outside of the dielectric sphere can be written
via the scalar function $\psi_{lm}(r,\theta,\phi)=\psi(r)Y_{lm}(\theta,\phi)$ where the radial solution is
\begin{equation}\label{radial}
\psi(r)=\left\{\begin{array}{cc} cj_l(\sqrt{\epsilon}k_0r) & \mbox{if $r<R$}\\
 aj_l(k_0r) +bh_l^{(1)}(k_0r)& \mbox{if $r\geq R$}
\end{array}\right.
\end{equation}
$j_l$ and $h_l^{(1)}$  are spherical Bessel and Hankel functions defined as
\begin{equation}\label{bessel}
j_l(x)=\sqrt{\frac{\pi}{2x}}J_{l+1/2}(x), ~~h_l^{(1)}(x)=\sqrt{\frac{\pi}{2x}}JH_{l+1/2}^{(1)}(x).
\end{equation}
$Y_{lm}$ are the spherical functions given by
\begin{eqnarray}\label{Ymn}
&Y_l^m(\theta,\phi)=(-1)^m\lambda_{lm}P_l^m(\cos\theta)e^{im\phi},\\
&P_l^m(x)=(1-x^2)^{m/2}\frac{1}{2l!}\frac{d^{l+m}}{dx^{l+m}}(x^2-1)^n,&\nonumber\\
&Y_l^{m^{*}}(\theta,\phi)=(-1)^mY_l^{-m}(\theta,\phi),&\nonumber
\end{eqnarray}
and
\begin{equation}\label{lambda}
    \lambda_{lm}=\sqrt{\frac{(2l+1)(l-m)!}{4\pi(l+m)!}}.
\end{equation}
Following Stratton \cite{Stratton} we introduce two independent vectorial fields expressed through a single
scalar function $\psi$ which satisfies the wave equation as follows
\begin{equation}\label{MN}
    \mathbf{M}_l^m=\nabla\times(\mathbf{r}\psi_{lm}),~~\mathbf{N}_l^m=\frac{1}{k}\nabla\times\mathbf{M}_l^m.
\end{equation}
Then for TE vector spherical modes we have
\begin{equation}
\label{MNTE}
 \left(\begin{array}{c} \mathbf{E}  \cr
 \mathbf{H}\end{array}\right)=\left(\begin{array}{c} \mathbf{M}_l^m  \cr
 -i\sqrt{\epsilon}\mathbf{N}_l^m\end{array}\right)
        \end{equation}
        and for the TM vector spherical modes
\begin{equation}
\label{MNTM}
 \left(\begin{array}{c} \mathbf{E}  \cr
 \mathbf{H}\end{array}\right)=\left(\begin{array}{c} \mathbf{N}_l^m  \cr
 -i\sqrt{\epsilon}\mathbf{M}_l^m\end{array}\right)
        \end{equation}

\section{Appendix B}
The value   $\mathcal{B}_{l\nu}^{00}$  is expressed via
\begin{equation}\label{B1}
\mathcal{H}(l,0,\nu,0,p)=\mathcal{G_{+}}+\mathcal{G_{-}}
\end{equation}
for $l+\nu+p$ odd
according to Eqs. (\ref{B})-(\ref{GG}) where
\begin{eqnarray}\label{B2}
&\mathcal{G_{\pm}}=\mp\sqrt{\nu(\nu+1)p(p-1)}\mathcal{G}(l,0,\nu,\pm 1,p-1)&\nonumber\\
&\mathcal{G}(l,0,\nu,\pm 1,p-1)=-\sqrt{(2l+1)(2\nu+1)(2p-1)}
\left(\begin{array}{ccc} l & \nu & p-1 \cr 0 & \pm 1 &\mp 1\end{array}\right)
\left(\begin{array}{ccc} l & \nu & p-1 \cr 0 & 0 & 0\end{array}\right)&
\end{eqnarray}
according to Eq. (\ref{G}).
Using the property of 3j-symbols
\begin{equation}\label{B3}
\left(\begin{array}{ccc} j_1 & j_2 & j_3 \cr m_1 & m_2 & m_3\end{array}\right)=
(-1)^{j_1+j_2+j_3}\left(\begin{array}{ccc} j_1 & j_2 & j_3 \cr -m_1 & -m_2 & -m_3\end{array}\right)
\end{equation}
we obtain
\begin{equation}\label{B4}
\left(\begin{array}{ccc} l & \nu & p-1 \cr 0 & 1 &-1\end{array}\right)=
\left(\begin{array}{ccc} l & \nu & p-1 \cr 0 & -1 & 1\end{array}\right)
\end{equation}
if $l+\nu+p-1$ is even.
Therefore we have from Eqs. (\ref{B1}) and (\ref{B2}) that $\mathcal{H}(l,0,\nu,0,p)=0$ and respectively,
$\mathcal{B}_{l\nu}^{00}=0$.
%%%%%%%%%%%%%%%%%%%%%%%%%%%%%%%%%%%%%%%%%%%%%%%%%%%

\end{document}